\documentclass[twocolumn,showpacs,preprintnumbers,amsmath,prl,amssymb,superscriptaddress]{revtex4}

\usepackage{graphicx}
\usepackage{dcolumn}
\usepackage{bm}
\usepackage{multirow}

\def\th232{\rm{ ^{232} Th }}
\def\u238{\rm{ ^{238} U }}
\def\cs137{\rm{^{137} Cs }}
\def\ba133{\rm{^{133} Ba }}
\def\cpd{\rm{kg^{-1}keV^{-1}day^{-1}}}
\def\nuebar{\rm{\bar{\nu_e}}}
\def\nue{\rm{\nu_e}}
\def\s2tw{\rm{ sin ^2 \theta _W }}
\def\munu{\rm{\mu _{\nu}}}
\def\munuebar{\rm{\mu _{\nuebar}}}
\def\mub{\rm{\mu _{B}}}
\def\enu{\rm{E_{\nu}}}
\def\rnusp{\rm{\phi ( \bar{\nu _e} ) }}
\def\ke10{\rm{\kappa _e}}

\def\halflife{\rm{\tau_{\frac{1}{2}}}}

\begin{document}

\preprint{AS-TEXONO/06-03}

\title{
Search of Neutrino Magnetic Moments with a
High-Purity Germanium Detector at 
the Kuo-Sheng Nuclear Power Station
}

%
\newcommand{\as}{Institute of Physics, Academia Sinica, Taipei 115, Taiwan.} 
\newcommand{\ntu}{Department of Physics, National Taiwan University,
Taipei 106, Taiwan.}
\newcommand{\ihep}{Institute of High Energy Physics, 
Chinese Academy of Science, Beijing 100039, China.}
\newcommand{\thu}{Department of Engineering Physics, Tsing Hua University,
Beijing 100084, China.}
\newcommand{\umd}{Department of Physics, University of Maryland,
College Park MD 20742, U.S.A.}
\newcommand{\ks}{Kuo-Sheng Nuclear Power Station, 
Taiwan Power Company, Kuo-Sheng 207, Taiwan.}
\newcommand{\iner}{Institute of Nuclear Energy Research, 
Lung-Tan 325, Taiwan.}
\newcommand{\ciae}{Department of Nuclear Physics, 
Institute of Atomic Energy, Beijing 102413, China.}
\newcommand{\bhu}{Department of Physics, Banaras Hindu University, 
Varanasi 221005, India.}
\newcommand{\metu}{Department of Physics, 
Middle East Technical University, Ankara 06531, Turkey.}
\newcommand{\nju}{Department of Physics, Nanjing University,
Nanjing 210093, China.}
\newcommand{\corr}{htwong@phys.sinica.edu.tw;
Tel:+886-2-2789-6789; FAX:+886-2-2788-9828.}

\affiliation{ \as }
\affiliation{ \ntu }
\affiliation{ \bhu }
\affiliation{ \umd }
\affiliation{ \metu }
\affiliation{ \ks }
\affiliation{ \iner }
\affiliation{ \ciae }
\affiliation{ \ihep }
\affiliation{ \thu }
\affiliation{ \nju }

\author{ H.T.~Wong } \altaffiliation[Corresponding Author: ]{ \corr } \affiliation{ \as }
\author{ H.B.~Li }  \affiliation{ \as } 
\author{ S.T.~Lin } \affiliation{ \as } \affiliation{ \ntu }
\author{ F.S.~Lee } \affiliation{ \as }
\author{ V.~Singh } \affiliation{ \as } \affiliation{ \bhu }
\author{ S.C.~Wu } \affiliation{ \as }
\author{ C.Y.~Chang } \affiliation{ \as } \affiliation{ \umd }
\author{ H.M.~Chang } \affiliation{ \as } \affiliation{ \ntu }
\author{ C.P.~Chen } \affiliation{ \as }
\author{ M.H.~Chou } \affiliation{ \as }
\author{ M.~Deniz } \affiliation{ \as } \affiliation{ \metu }
\author{ J.M.~Fang } \affiliation{ \ks }
\author{ C.H.~Hu } \affiliation{ \iner }
\author{ H.X.~Huang } \affiliation{ \ciae }
\author{ G.C.~Jon } \affiliation{ \as }
\author{ W.S.~Kuo } \affiliation{ \iner }
\author{ W.P.~Lai } \affiliation{ \as } 
\author{ S.C.~Lee } \affiliation{ \as }
\author{ J.~Li }  \affiliation{ \as } \affiliation{ \ihep } \affiliation{ \thu }
\author{ H.Y.~Liao } \affiliation{ \as } \affiliation{ \ntu }
\author{ F.K.~Lin } \affiliation{ \as }
\author{ S.K.~Lin } \affiliation{ \as }
\author{ J.Q.~Lu } \affiliation{ \as } \affiliation{ \nju }
\author{ H.Y.~Sheng } \affiliation{ \as } \affiliation{ \ihep }
\author{ R.F.~Su } \affiliation{ \ks }
\author{ W.S.~Tong } \affiliation{ \iner }
\author{ B.~Xin } \affiliation{ \as } \affiliation{ \ciae }
\author{ T.R.~Yeh } \affiliation{ \iner }
\author{ Q.~Yue } \affiliation{ \thu }
\author{ Z.Y.~Zhou } \affiliation{ \ciae }
\author{ B.A.~Zhuang } \affiliation{ \as } \affiliation{ \ihep }

\collaboration{TEXONO Collaboration}

\noaffiliation


\date{\today}

\begin{abstract}

A search of neutrino magnetic moments
was carried out 
at the Kuo-Sheng Nuclear Power Station
at a distance of
28~m from the 2.9~GW reactor core.
With a high purity germanium
detector of mass 1.06~kg surrounded by 
scintillating NaI(Tl) and CsI(Tl) crystals as
anti-Compton detectors, 
a detection threshold of 5~keV and a background level
of 1~$\cpd$ near threshold were achieved.
Details of the reactor neutrino source,
experimental hardware, 
background understanding 
and analysis methods are presented.
Based on 570.7 and 127.8 days of 
Reactor ON and OFF data, respectively,
at an average Reactor ON
electron anti-neutrino flux
of $\rm{6.4 \times 10^{12} ~ cm^{-2} s^{-1} }$,
the limit on the neutrino magnetic moments of
$\rm{ \munuebar < 7.4 \times 10^{-11} ~ \mub}$
at 90\%  confidence level was derived.
Indirect bounds on the $\nuebar$ radiative decay
lifetimes were inferred.
\end{abstract}

\pacs{14.60.Lm, 13.15.+g, 13.40.Em}

\maketitle

\section{I. Introduction}

The strong evidence of neutrino oscillations
from the solar, atmospheric as well as long baseline
accelerator and reactor neutrino measurements
implies finite neutrino masses and mixings~\cite{pdg04,nu04}.
Their physical origin and experimental consequences
are not fully understood.
Experimental studies on the neutrino properties
and interactions can shed light to these
fundamental questions and provide constraints to
the interpretations in the
future precision oscillation experiments.

The couplings of neutrinos with the photons are
generic consequences of finite
neutrino masses, and are one of the
important intrinsic neutrino properties~\cite{nuprop}
to explore.
The neutrino electromagnetic vertex can be
parametrized by terms 
corresponding to interactions without and with
its spin, identified as
the ``neutrino charge radius'' and
``neutrino magnetic moments'', respectively.

This article reports on a search of the neutrino
magnetic moments with reactor neutrinos at the
Kuo-Sheng (KS) Nuclear Power Station in Taiwan.
It extends over our previous publication~\cite{prl03}
and covers the 
reactor neutrino spectra ($\rnusp$)
and the other experimental features in details.
A factor of three larger data sample was used, 
and an analysis procedure combining
information from all measured spectra was devised.

\section{II. Neutrino Magnetic Moments}

\begin{figure}[hbt]
\centerline{
\includegraphics[width=8.0cm]{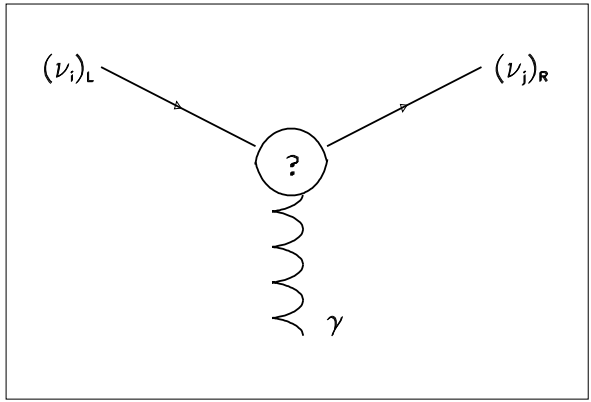}
}
\caption{
The schematic diagram of the
neutrino-photon interaction
vertex which involves a change
of the helicity-states. The
coupling is parametrized by the
neutrino magnetic moment.
}
\label{vertex}
\end{figure}

An overview on the particle physics aspects
of neutrino magnetic moments can be referred to a 
recent review~\cite{munureview} and the
references therein. The neutrino 
magnetic moment ($\munu$) is 
an experimentally observable parameter 
which characterizes a possible
coupling between neutrino mass
eigenstates $\nu _i$ and
$\nu _j$ with the photon whereby
the helicity-state is flipped: 
\begin{equation}
( \nu _i ) _L ~ - ~ \gamma ~ - 
( \nu _j ) _R 
\end{equation}
as depicted
schematically in Figure~\ref{vertex}.
The vertex marked ``?'' denotes
the unknown physics to be explored.
The parameter $\munu$ is usually expressed in 
units of the Bohr magneton:
\begin{equation}
\mub =  \frac{e}{2 m _e}  ~~ ; ~~ e^2 = 4 \pi \alpha _{em} ~~ ,
\end{equation}
where $\alpha _{em}$ is the fine-structure constant
and $m _e$ is the electron mass.
Both {\it diagonal} and {\it transition}
moments are possible, corresponding to the cases where
$ i = j$ and $i \neq j$, respectively.
Symmetry principles place constraints on the
possible channels~\cite{kaysernieves}
and require that the diagonal moments vanish
for Majorana neutrinos.
The study of neutrino magnetic moments
is, in principle, a way to distinguish
between Dirac and Majorana neutrinos $-$
a crucial unresolved issue in neutrino physics.
For example, if a finite $\munu$ is observed in the laboratory,
and electron anti-neutrinos ($\nuebar$) from the Sun is detected
with a spectrum 
consistent with the oscillation and solar model
parameters, neutrinos would be
Majorana particles.
Recent derivations~\cite{munubounds} 
of model-independent $\munu$-ranges
indicated that upper bounds for Dirac neutrinos 
are several orders of magnitude more stringent than the
current experimental limits. 
Consequently, observations
of $\munu$ at the present sensitivities will
imply that neutrinos are Majorana. 

Once a model is specified,
the vertex ``?'' is known
and $\munu$ can be calculated
from first principles.
Minimally-Extended Standard Model with massive
Dirac neutrinos~\cite{mesm} gives
$\munu  \sim 10^{-19}  [ m_{\nu} / 1~\rm{eV} ] ~ \mub$
which is far too small to have any observable
consequences. Incorporation
of additional physics, such as
Majorana neutrino transition moments or
right-handed weak currents, can significantly
enhance $\munu$ to the experimentally relevant
ranges~\cite{bsm}.
Supersymmetry as well as extra-dimensions~\cite{extradim}
can also contribute
to the process.

The parameter $\munu$ for
neutrinos with energy $E_{\nu}$ produced as
$\nu_l$ at the source and after traversing a distance L
is given by~\cite{vogelbeacom}
\begin{equation}
\munu^2 ( \nu_{\it l} , L , E_{\nu} ) =
\sum_{j} | \sum_{i} U_{\it li} ~ e^{-i E_{\nu} L } ~ \mu_{ij} | ^2  ~~ ,
\end{equation}
where U$_{li}$ is the neutrino mixing matrix
and $\mu_{ij}$ are the coupling constants
between 
$\rm{\nu_i}$ and $\rm{\nu_j}$ with the photon.
Consequently, $\munu$  is
therefore
an effective and convoluted parameter
and the interpretations of
experimental results depend on
the exact $\nu_l$ compositions at
the detectors.
In this paper, we write
$\rm{\munuebar = \ke10 \times 10^{-10} ~ \mub}$
and work with the parameter $\ke10$
for simplicity.

The neutrino spin-flavor
precession (SFP) mechanism~\cite{sfp}, 
with or without matter
resonance effects in the solar medium,
has been used to explain solar neutrino
deficit~\cite{sfpsolar}.  This scenario
is compatible with all solar neutrino data.
The terrestrial KamLAND experiment, however,
recently confirmed the Large Mixing
Angle (LMA) parameter space
of the matter oscillation
scenario as {\it the} solution
for the solar neutrino problem~\cite{kamland}, such
that SFP can be excluded as
the dominant contribution in 
solar neutrino physics~\cite{munusolar}.
Alternatively, the measured solar
neutrino $\nu_{\odot}$-e spectral shape
has been used to set limit of $\rm{\kappa_{\odot} < 1.1}$
at 90\% confidence level (CL) for the  ``effective'' $\nu_\odot$
magnetic moment~\cite{vogelbeacom,skmunu}
which is different
from that of a pure $\nuebar$ state derived in reactor experiments.
Other astrophysical bounds on $\munu$ were mostly
derived from the consequences from
a change of the neutrino spin-states
in the astrophysical medium\cite{munureview,raffeltbook},
typically of the range
$\rm{\kappa_{astro} < 10^{-2} - 10^{-3}}$.
However, these bounds are model-dependent
and involve implicit assumptions
on the neutrino properties.

Direct laboratory experiments~\cite{munureview}
on $\munu$
utilize  accelerator and reactor neutrinos
as sources, and are conducted under
controlled conditions.
The most sensitive searches 
are usually performed by experiments studying
neutrino-electron scatterings~\cite{vogelengel}:
\begin{equation}
  \nu_{\it l}   +   e^-   
\rightarrow  
  \nu_{\it x}   +   e^-  .
\end{equation}
The experimental observable is the kinetic energy of the
recoil electrons (T).
A finite $\munu$
will contribute to a differential cross-section term
given by:
\begin{equation}
\label{eq::mm}
( \frac{ d \sigma }{ dT } ) _{\munu}  ~ = ~
\frac{ \pi \alpha _{em} ^2 {\it \munu } ^2 }{ m_e^2 }
 [ \frac{ 1 - T/E_{\nu} }{T} ] ~ .
\end{equation}
The signature is
an excess of events
over those due to Standard Model
of electroweak interactions~(SM)
and other background processes, which
exhibit the characteristic 1/T spectral dependence.
Limits from negative searches
are valid for both Dirac and Majorana
neutrinos and for both
diagonal and transitional moments.

The neutrino radiative decay~\cite{rdk}
for the process 
\begin{equation}
\nu_i ~ \rightarrow ~ \nu_j ~ + ~ \gamma
\end{equation}
is another manifestation of the
neutrino electromagnetic couplings
where a change of the
neutrino helicity-states takes place.
A final-state real photon is produced
in the process,
unlike the $\munu$-effects on
neutrino-electron scatterings, where
only virtual photons are involved.
The decay rate $\Gamma _{ij}$ and
the decay lifetime $\tau _{ij}$
is related to $\mu_{ij}$ via~\cite{rdkmunu}
\begin{equation} 
\label{eq::rdk}
\Gamma_{ij} = \frac{ 1 }{ \tau _{ij} } =
\frac{1}{8 \pi} \frac{( m_i^2 - m_j^2 ) ^ 3}{m_i^3}
\mu_{ij}^2  ~ ~ ,
\end{equation}
where $m_{i,j}$ are the masses for
$\nu_{i,j}$.

\section{III. Direct Searches with Reactor Neutrinos}

\begin{figure}[hbt]
\centerline{
\includegraphics[width=8.0cm]{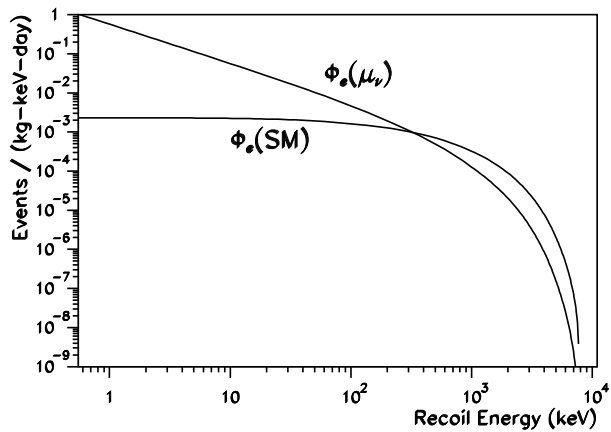}
}
\caption{
The expected recoil spectra due to
neutrino-electron scattering with
reactor $\nuebar$ at a flux of
$\rm{10^{13}~cm^{-2} s^{-1}}$.
Contributions from 
Standard Model [$\Phi _e ( SM )$]
and magnetic moments [$\Phi _e ( \munu )$]
at $\rm{10^{-10} ~ \mub}$ are shown.
}
\label{recoil}
\end{figure}

Reactor neutrinos provide a sensitive probe
for laboratory searches  of $\rm{\munuebar}$,
taking advantages of the
high $\nuebar$ flux, low $\enu$ and better experimental
control via the reactor ON/OFF comparison.
The measurable electron recoil energy spectra due to 
SM and $\munu$ (at $\ke10$=1.0) 
scatterings with reactor-$\nuebar$'s,
denoted respectively
by $\Phi _e ( SM )$ and $\Phi _e ( \munu )$,
are displayed in Figure~\ref{recoil}. 
A finite $\rm{\munuebar}$ would manifest itself
as excess of events in the ON spectra over the
background as derived from the OFF data,
which have a 1/T energy spectrum.

Neutrino-electron scatterings were first
observed in the pioneering experiment~\cite{reines}
at Savannah River where
plastic scintillators were adopted as target
surrounded by NaI(Tl) crystal scintillators as
anti-coincidence counters.
A revised analysis of the data by Ref~\cite{vogelengel}
with improved input parameters
gave a positive signature
consistent with the interpretation of a
finite $\munuebar$ at $\ke10$=2$-$4.
An intrinsic error source of this experiment was
the use of a proton-rich target.
The cross-section for the proton-capture
reaction:
\begin{equation}
\label{eq::nuebarproton}
\nuebar + p \rightarrow e^+ + n
\end{equation}
is much larger than that for $\nuebar$-e
scatterings.
Other results came
from the Kurtchatov~\cite{kurt} and Rovno~\cite{rovno} experiments
which quoted limits of $\ke10 < 2.4$
and $< 1.9$ at 90\% CL, respectively.
A recent experiment MUNU~\cite{munu}
at the Bugey reactor adopted
a time projection chamber with CF$_4$ gas
surrounded by active liquid scintillator 
as anti-Compton vetos.
A limit of  $\ke10 < 0.9$ at 90\% CL
was set with 66.6~days of data at
a threshold of 700~keV.
The lowest-energy bin was at
a $\sim 2 \sigma$ excess over the SM value.

A global analysis was performed~\cite{global}
which combined simultaneously
the $\munu$ data from the
reactor and solar neutrino experiments,
as well as the
LMA oscillation parameters constrained by
solar neutrino and KamLAND results.
Only Majorana neutrinos were considered such
that there were only transition moments.
A ``total'' magnetic moment vector
$\Lambda$=$( \mu_{23} , \mu_{31} , \mu_{12} )$
was defined, such that its
amplitude was given by
$ | \Lambda | ^ 2$=$\frac{1}{2} Tr ( \mu ^{+} \mu ) $.
A global fit produced 90\% CL limits of
$  | \Lambda |  <  4.0 \times 10^{-10} ~ \mub$
from solar and KamLAND data only,
while
$  | \Lambda |  <  1.8 \times 10^{-10} ~ \mub$
when reactor data were added.
The results indicate the important
role of
reactor experiments in constraining
the magnetic moment effects.

The systematic effects related to the poorly-known
neutrino spectra were difficult to control
in the 1~MeV range relevant to the 
previous experiments~\cite{lenu}.
The approach of the KS experiment is to achieve
a threshold of $\sim$10~keV range through a
matured and reliable detector technology $-$
high-purity germanium (HPGe) detector.
Three important advantages can be realized which
significantly enhance the sensitivities and
robustness of the results:
(1) the potential signal rate is much increased due
to the 1/T energy dependence of Eq.~\ref{eq::mm};
(2) as shown in Figure~\ref{recoil},
the event rates from $\Phi _e ( \munu )$ 
are 24~times those of $\Phi _e ( SM )$ 
at T$\sim$10~keV and $\ke10$$\sim$1, 
such that
the uncertainties in the SM background do not affect
the $\munu$ signals;
(3) Eq.~\ref{eq::mm} is mostly
independent of $\enu$ at 
$\sim$10-100~keV, such that the $\munu$ signals
depends only on the well-known total reactor neutrino flux
but {\it not} the details of $\rnusp$,
thereby reducing the systematic uncertainties.
In contrast, 
the poorly-modeled $\rnusp$ at $\sim$1~MeV 
gives rise to uncertainties
in the irreducible $\Phi _e ( SM )$ background 
(which is 2.7~times 
the $\Phi _e ( \munu )$-signals at
$\ke10 \sim 1$)
for experiments operating at this energy range.

We report in this article data taken at KS with HPGe.
There are three Reactor ON/OFF 
data taking periods, the key information
of which are summarized in Table~\ref{daqperiod}.

\begin{table*}[hbt]
\caption{\label{daqperiod}
Summary of the key information on the three data taking periods.
}
\begin{ruledtabular}
\begin{tabular}{lccccccc}
Period & Data Taking  & Reactor ON &
Reactor ON  & Reactor OFF & Reactor OFF & DAQ &
Average \\
& Calendar Time & Real Time & Live Time &
Real Time & Live Time &  Live Time &  $\nuebar$ flux \\
& & (days) & (days) & (days) & (days) & (\%)
& ($10^{12} ~ \rm{ cm^{-2} s^{-1} }$) \\ \hline
I & July 2001 - April 2002 &  188.2 & 180.1 & 55.1 & 52.7 & 95.7 & 6.29 \\
II & Sept. 2002 - April 2003 &  125.8 & 111.7 & 34.4 & 31.5 & 89.4 & 6.53 \\
III & Sept. 2004 - Oct. 2005 &  303.9 & 278.9 & 48.7 & 43.6 & 91.5 & 6.51 \\ \hline
Total & $-$ &  617.9 & 570.7 & 138.2 & 127.8 & 92.4 & 6.44
\end{tabular}
\end{ruledtabular}
\end{table*}

\section{IV. Reactor Neutrino Spectrum}

The $\nuebar$'s emitted in
power reactors
are predominantly
produced through $\beta$-decays of
(a) the fission products, following the
fission of the four dominant fissile isotopes:
$^{235}$U, $^{238}$U, $^{239}$Pu and $^{241}$Pu,
and
(b) $^{239}$U, following
the neutron capture on the $^{238}$U fuel:
$^{238}$U(n,$\gamma$)$^{239}$U~.

\begin{table*}[hbt]
\caption{\label{rnu}
Fractional
mass compositions, relative fission contributions
and neutrino yield per fission for the four fissile
isotopes and the $^{238}$U neutron capture.
All values are time-dependent and only typical
levels are given.
}
\begin{ruledtabular}
\begin{tabular}{lcccc}
Channels & Fractional Compositions
& Relative Rates & Neutrino Yield  & Neutrino Yield \\
& by Mass (\%) & per Fission & per Event & per Fission \\ \hline
$^{235}$U Fission & 1.5 & 0.55  &  6.14 & 3.4 \\
$^{238}$U Fission & 98.0 & 0.07  &  7.08 & 0.5 \\
$^{239}$Pu Fission & 0.4 & 0.32  &  5.58 & 1.8 \\
$^{241}$Pu Fission & $<$0.1 & 0.06  &  6.42 & 0.4 \\
$^{238}$U (n,$\gamma$) $^{239}$U & $-$ & 0.60 & 2.00 & 1.2 \\ \hline
Total & $-$ & $-$ & $-$ & 7.3
\end{tabular}
\end{ruledtabular}
\end{table*}

\begin{figure}[hbt]
\centerline{
\includegraphics[width=8.0cm]{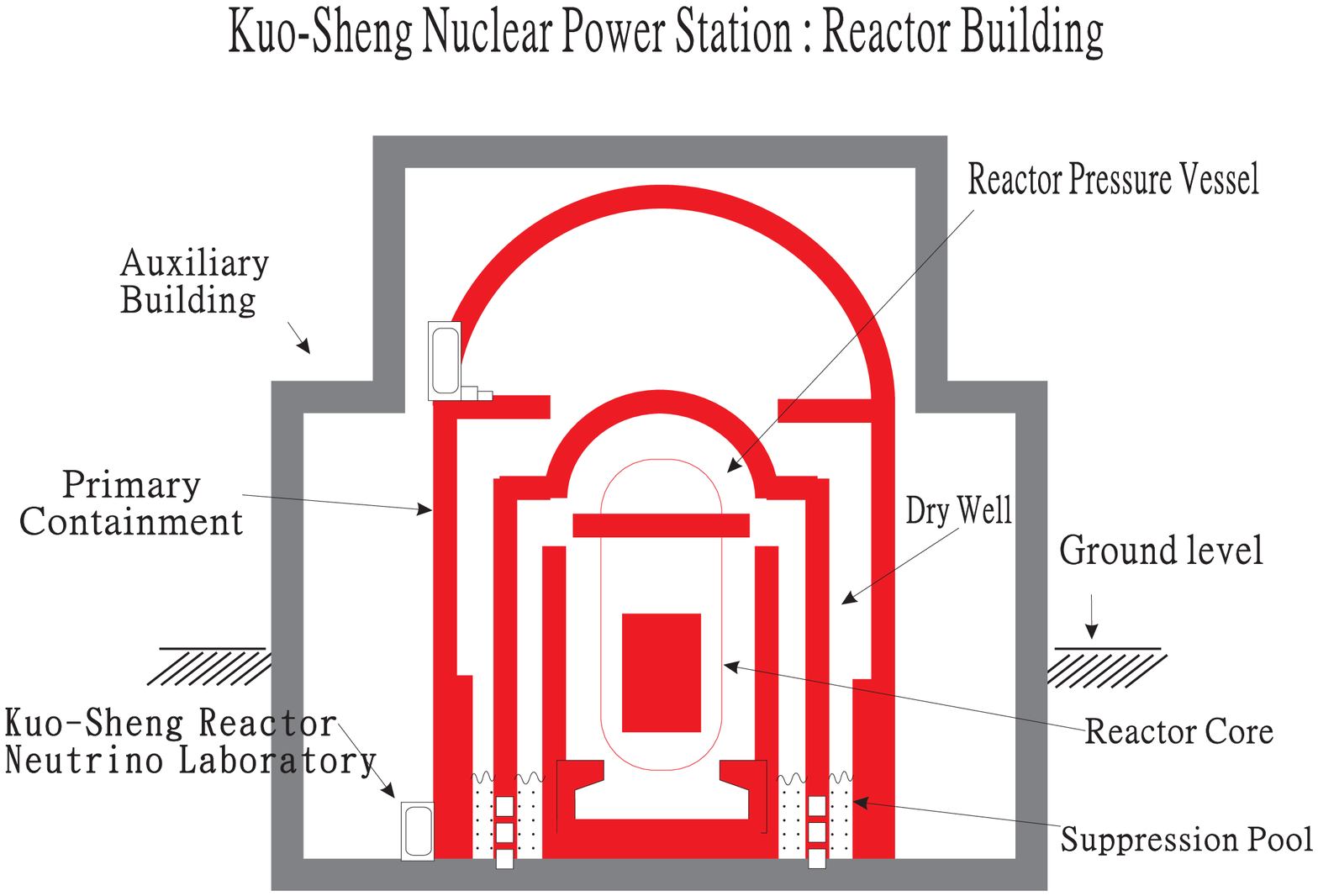}
}
\caption{
Schematic layout of the Kuo-Sheng
Neutrino Laboratory in relation to the
core of the power reactor. 
}
\label{kslab}
\end{figure}

\begin{figure}[hbt]
{\bf (a)}\\
\includegraphics[width=8.0cm]{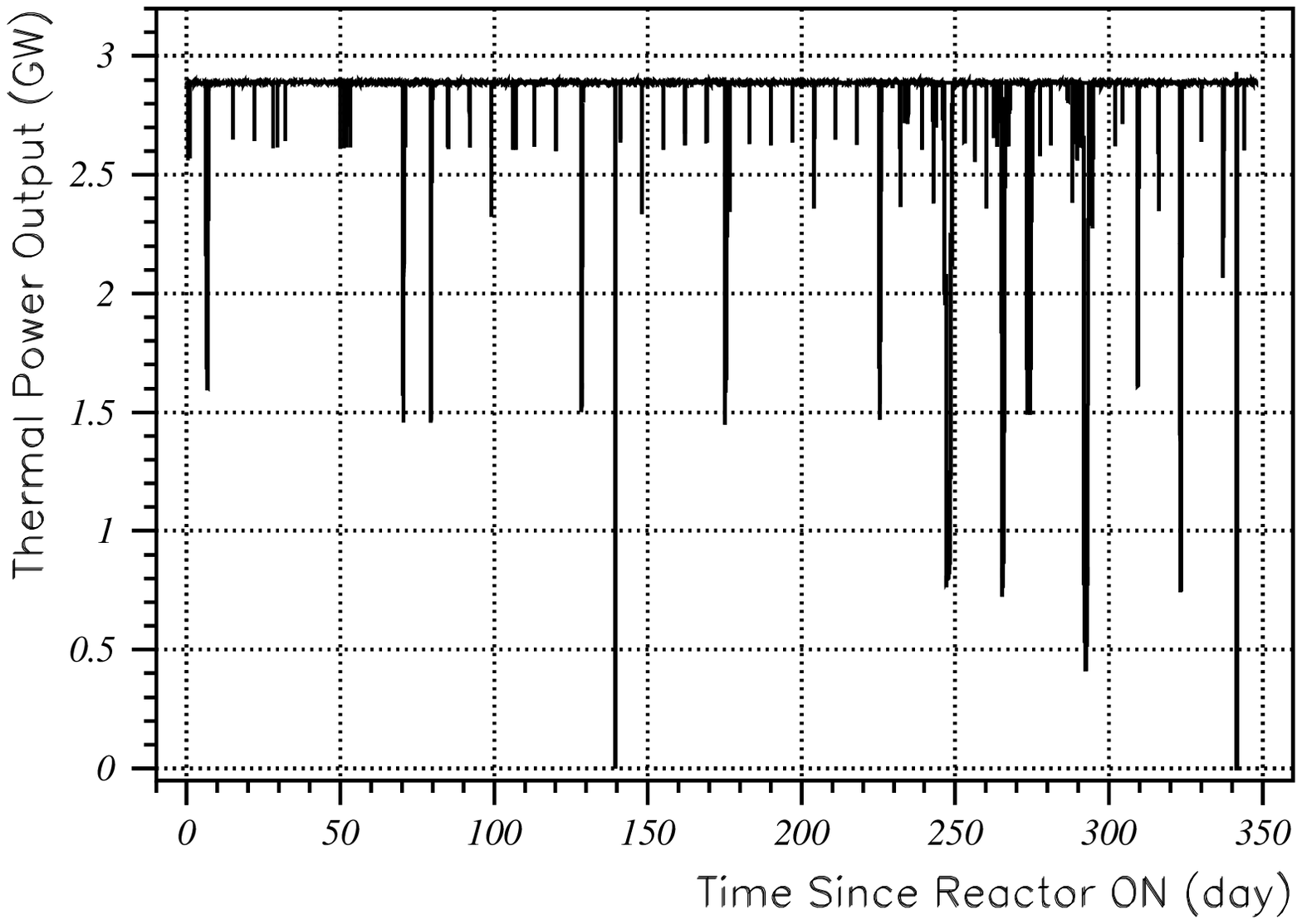}\\
{\bf (b)}\\
\includegraphics[width=8.0cm]{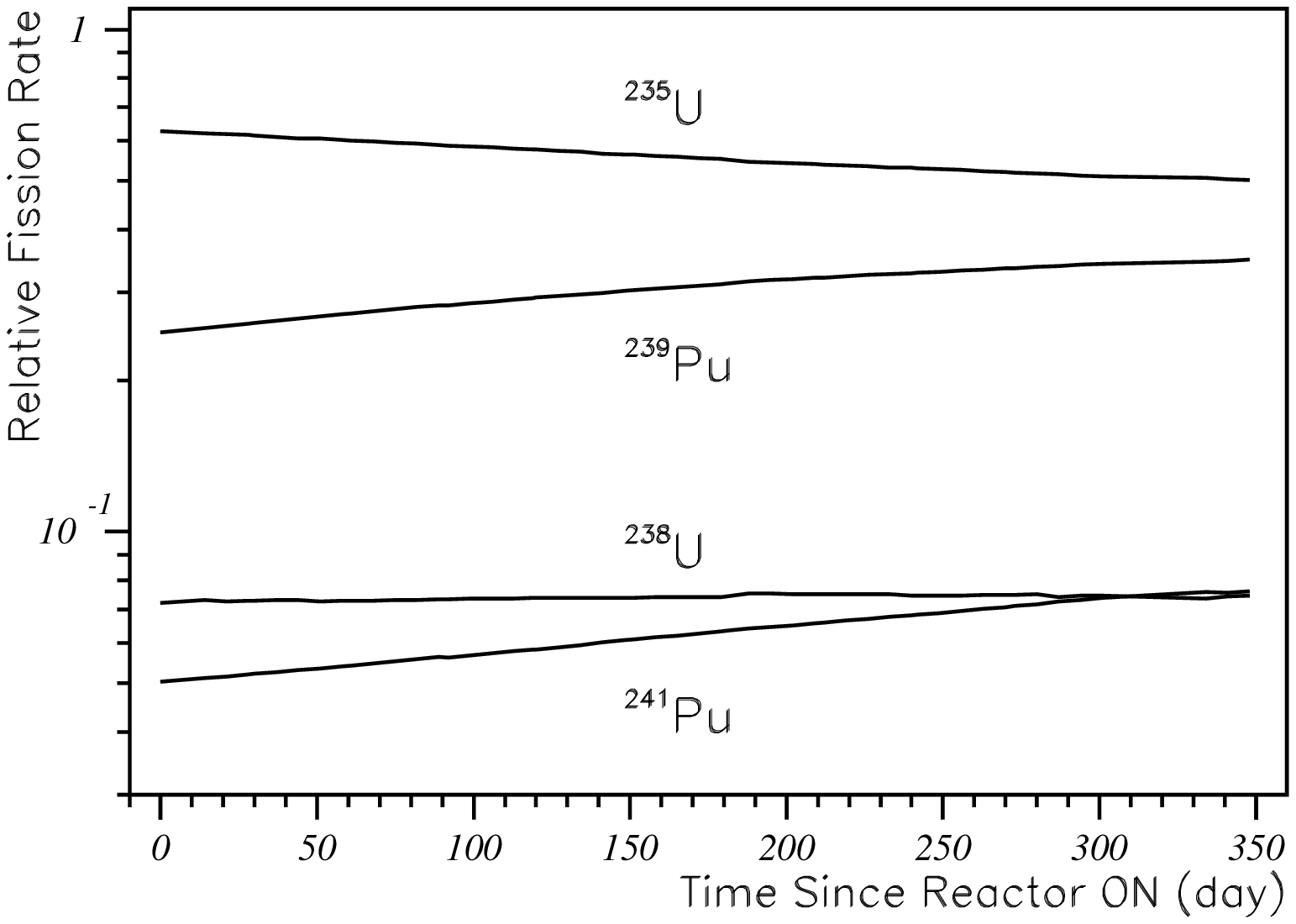}\\
{\bf (c)}\\
\includegraphics[width=8.0cm]{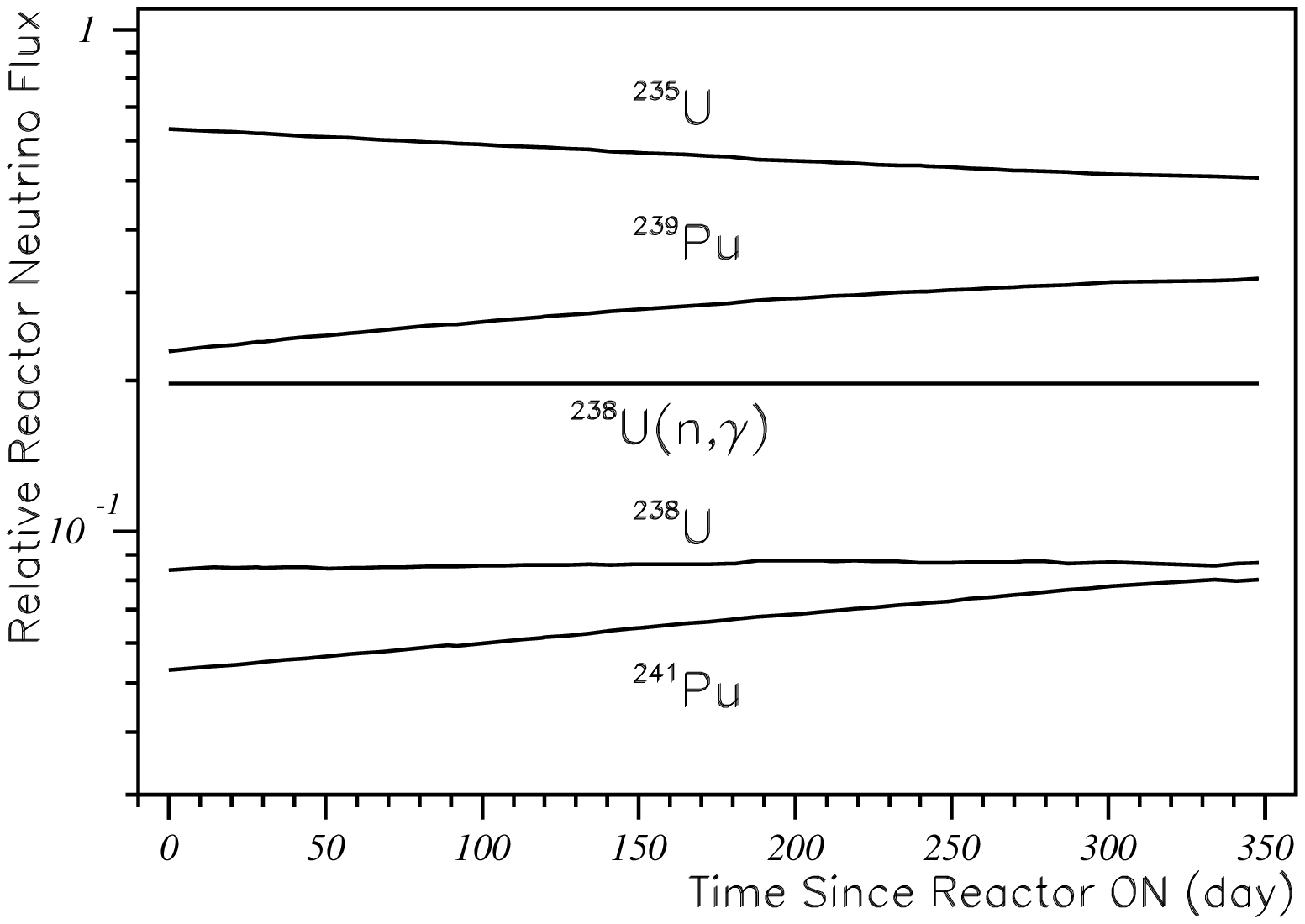}
\caption{
The variations of the (a) thermal power output,
as well as the (b) relative fission
rates and (c) $\nuebar$-flux 
of the fissile isotopes over Period-III
of data taking. 
The reactor outage OFF period
was completed on Day-0.
}
\label{reactordata}
\end{figure}

The KS neutrino laboratory,
as depicted schematically in Figure~\ref{kslab},
is located at a distance
of 28~m from 
Core \#1 of the Kuo-Sheng Nuclear Power
Station in Taiwan.
The nominal thermal power output is 2.9~GW. 
The standard operation includes 
about 18~months
of Reactor ON time at nominal power
followed by about 50~days of 
Reactor outage OFF period when about
a third of the fuel elements are replaced.
Reactor operation data on the thermal power output 
and control rod status
as functions of time and locations within the core
were provided to the experiment by the
Power Station. 
A set of software programs~\cite{inersoftware} 
was specifically developed, in
association with the 
commercial SIMULATE-3 
and CASMO-3 codes~\cite{comsoftware}, 
both of which are extensively used in the field 
of nuclear reactor core analyses.
The variations of the thermal power output,
as well as
the fission rates and $\nuebar$-flux 
of the fissile isotopes 
during Period-III 
are displayed in Figures~\ref{reactordata}a-c.
At steady state operation, the total fission rates 
and the total neutrino fluxes were 
constant to better than 0.1\% and 0.2\%, respectively.
Data taken during the short durations of 
unscheduled reactor stops
were included into the Reactor-OFF category.

The typical $\nuebar$-yield for these channels,
as well as their
relative contributions per fission
are summarized in Table~\ref{rnu}.
The $\nuebar$-yields for the fission elements
were adopted from Ref.~\cite{vogel81}, following
a survey of the fission daughter 
isotopes and the subsequent $\beta$-decays necessary
to reach stability. 
The $^{238}$U neutron capture rate of 0.6 per fission were 
evaluated by two independent methods:
(a) via full neutron transport calculations of the neutrons
in the reactor core~\cite{ncaprussian,rnue}, and 
(b) by evaluating the difference 
in the decrease of the amount of $^{238}$U 
and the $^{238}$U fission rate. 
Results derived
by both methods were consistent to
a few percent.



\begin{figure}[hbt]
{\bf (a)}\\
\includegraphics[width=8.0cm]{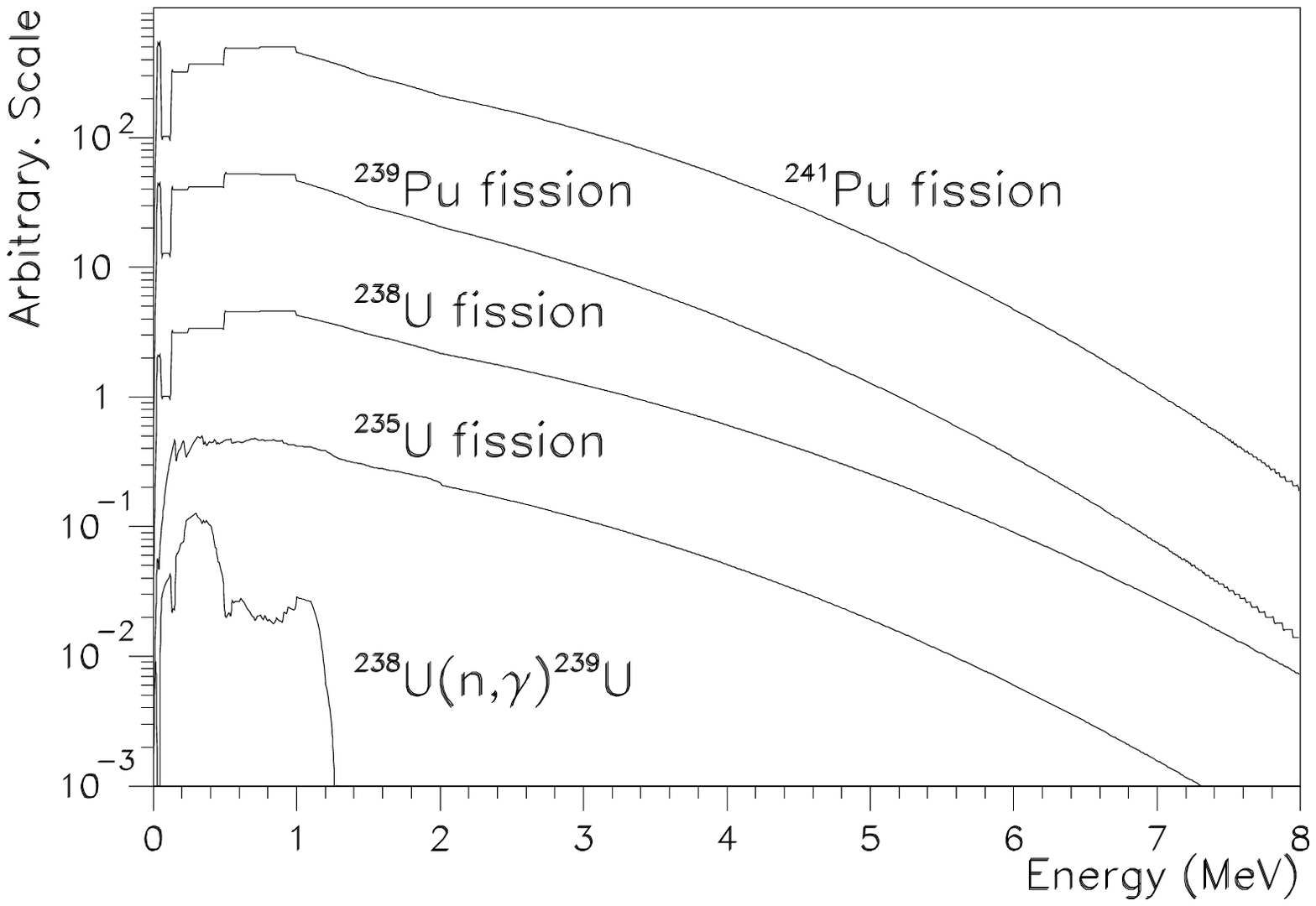}\\
{\bf (b)}\\
\includegraphics[width=8.0cm]{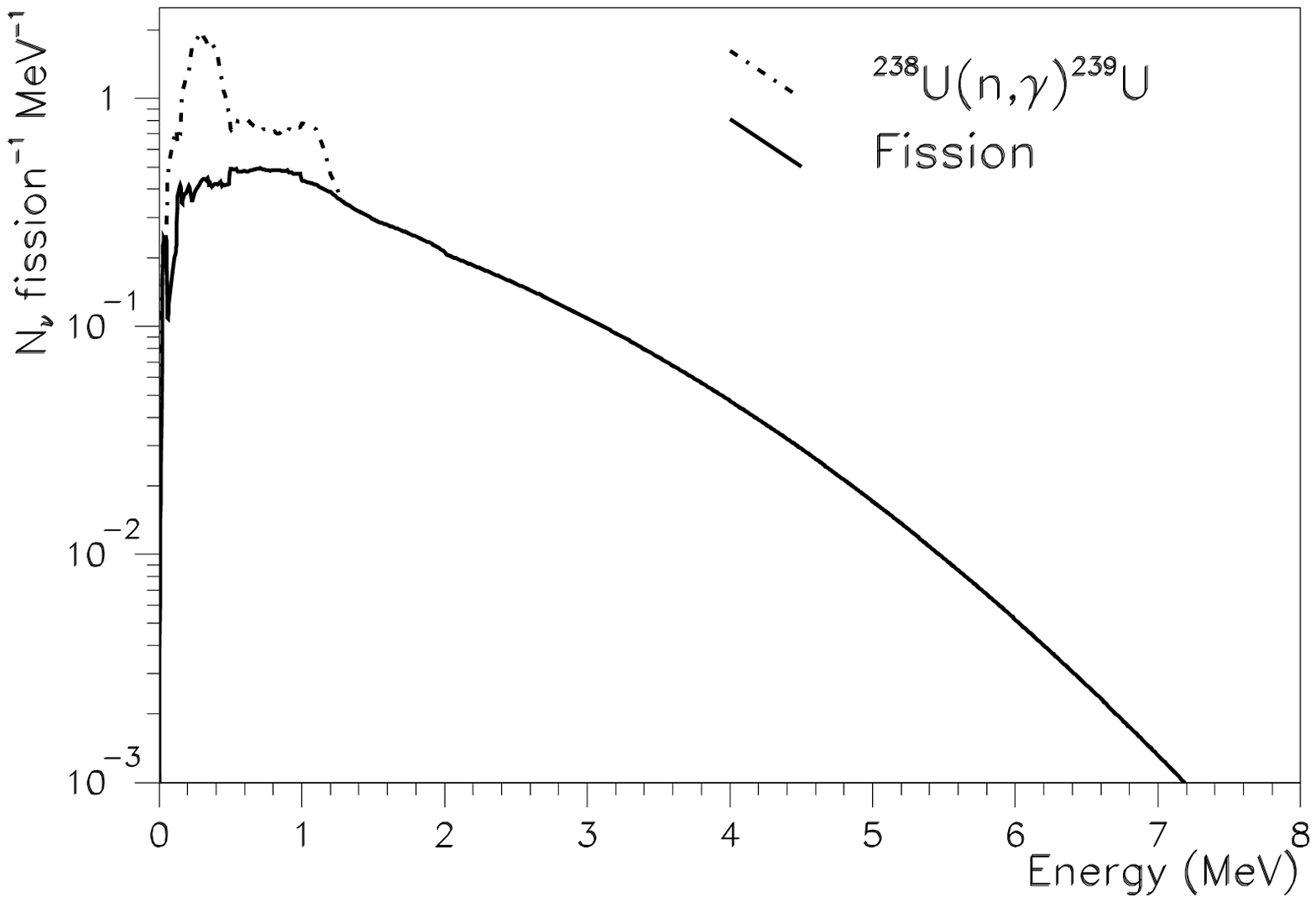}
\caption{
(a)
Spectral shape for reactor neutrinos
due to
individual production channels.
(b)
Total spectrum at the typical power
reactor operation.
}
\label{rnusp}
\end{figure}

The $\rnusp$'s of the five
channels, as depicted in Figure~\ref{rnusp}a,
were adopted from Ref.~\cite{vogelengel} for the
fission $\nuebar$'s, while those following
 $^{238}$U neutron capture were derived from
standard $\beta$-spectra of $^{239}$U and
$^{240}$Np.
The components were summed according
to the relative contributions per fission, and
the resulting total $\rnusp$
is shown in Figure~\ref{rnusp}b.
This spectrum was used as input in deriving
the expected
electron recoil spectra in Figure~\ref{recoil}.
The evaluated $\nuebar$-fluxes for the three
periods are given
in Table~\ref{daqperiod}, where the weighted
average is
$\rm{6.4 \times 10^{12} ~ cm^{-2} s^{-1} }$.

The modeling of $\rnusp$
above the threshold of the reaction
of Eq.~\ref{eq::nuebarproton} from
reactor operation data
has been well-established.
Accuracies of better than 1.4\% and 5\%
between calculations and measurements
were achieved in the
integrated~\cite{bintflux}
and differential~\cite{bugey3} spectra.
The evaluation of $\rnusp$ 
at energy below 2~MeV, on the other hand,
is much more complicated~\cite{lenu}.
Many input parameters remain unknown 
and there are no
measurements to cross-check.
Consequently, the 
$\Phi _e ( SM )$ recoil spectra 
below the MeV range
in Figure~\ref{recoil} were subjected
to large uncertainties.
However,  the spectra
$\Phi _e ( \munu )$ at
10-100~keV were accurately predicted due to
the $\enu$-independence of Eq.~\ref{eq::mm}. 
They followed a 1/T profile 
and depended only on 
the total $\nuebar$-fluxes of
Table~\ref{daqperiod}, which were calculated 
with an expected accuracy of a few percent.

\section{V. Experimental Set-Up}

\begin{figure}[hbt]
\centerline{
\includegraphics[width=8.0cm]{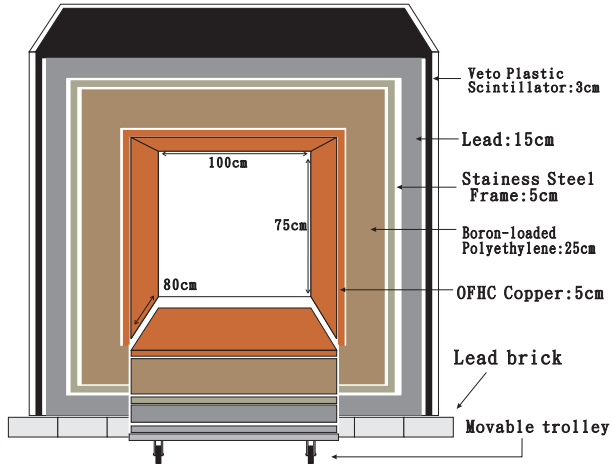}
}
\caption{
The shielding design of the KS Neutrino
Laboratory. Detectors and inner shieldings were
placed in the inner target volume. 
}
\label{shielding}
\end{figure}

A research program
on low energy neutrino physics~\cite{ksprogram}
is being pursued by the TEXONO
Collaboration at the KS Neutrino Laboratory.
The laboratory is equipped
with an outer 50-ton shielding structure
depicted schematically in Figure~\ref{shielding},
consisting of, from outside in,
2.5~cm thick plastic scintillator panels with
photo-multiplier tubes (PMTs) readout
for cosmic-ray veto (CRV),
15~cm of lead, 5~cm of stainless
steel support structures, 25~cm of boron-loaded polyethylene
and 5~cm of OFHC copper.
The innermost volume with a dimension of
100$\times$80$\times$75~$\rm{cm^3}$
provides the flexibilities of placing different
detectors for different physics topics.
During the I-III data taking periods listed in 
Table~\ref{daqperiod}, 
both the HPGe and a CsI(Tl) scintillating
crystal array~\cite{kscsi} together with
their associated inner shieldings were placed
in the inner volume. The CsI(Tl) array is
for the measurement of neutrino-electron
scattering cross-sections.
The $\munu$-search
reported in this article
was performed with the HPGe detector.

As indicated in Figure~\ref{kslab},
the laboratory is located at
the ground floor of the reactor building
at a depth of 12~m below sea-level and with
about 25~m water equivalence of overburden.
The primary cosmic-ray hadronic components
are eliminated while the muon flux is reduced
by a factor of 4.
Ambient $\gamma$-background at the reactor site
is about 10 times higher in the MeV range
than that of a typical laboratory,
dominated by activity due to 
$^{60}$Co and $^{54}$Mn (half-lives
5.27~years and 312~days, respectively)
present as dust in the environment.
Both isotopes are produced 
by neutron activation on the construction
materials at the reactor core.
The dust can get settled on exposed surfaces
within hours and is difficult to remove.
Some earlier prototype detectors were contaminated
by such. Attempts to clean the surfaces in situ
resulted in higher contaminations.
Accordingly, the various detector and
inner shielding components
were carefully washed and wrapped by 
several layers of plastic sheets 
before transportation to KS. 
Outer layers were removed in situ 
only prior to installation, while
inner layers were replaced between 
data taking periods.
Neutron background is comparable to that of a typical
surface location, with no observable differences between the
ON and OFF periods.

\begin{figure}[hbt]
\includegraphics[width=8cm]{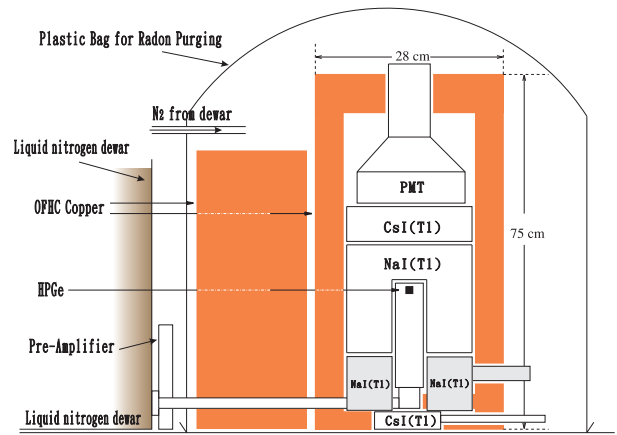}
\caption{
Schematic layout of the HPGe
with its anti-Compton detectors
as well as inner shieldings and
radon purge system. 
}
\label{kshpge}
\end{figure}

The HPGe set-up is shown schematically in
Figure~\ref{kshpge}.
It is a coaxial germanium detector~\cite{canberra} 
with an active target mass of 1.06~kg. 
The construction materials and detector geometry
followed the ``Ultra Low Background (ULB)''
standards.
The lithium-diffused outer
electrode is 0.7~mm thick. The end-cap cryostat, also
0.7~mm thick, is made of OFHC copper. Both of these
features provide total suppression to 
ambient $\gamma$-background
below 60~keV, such that events below this energy are either
due to internal activity or ambient MeV-range $\gamma$'s
via Compton scattering.
Consequently, the background profile should be smooth
and continuous below 60~keV down to the ranges where
atomic effects are important.
The HPGe was surrounded by an 
anti-Compton veto (ACV) detector system
made up of three components: 
(1) an NaI(Tl) ``well-detector''
of thickness 5~cm that fit onto the end-cap cryostat and
is coupled to a 12~cm PMT readout through an additional
7~cm of CsI(Tl) as active light guide;
(2) an NaI(Tl) ``ring-detector'' of thickness 5~cm 
at the joint of the cryostat; 
and (3) a 4~cm thick CsI(Tl) ``base-detector'' at the bottom.
All of these ACV detectors were
assembled within mechanical structures
made of OFHC copper, and 
were read out by PMTs with low-activity glass.
The assembly was further surrounded by 3.7~cm of 
OFHC copper inner shielding. 
Another OFHC copper wall 
of thickness 10~cm provided 
additional shieldings on the side of 
the liquid nitrogen dewar and
pre-amplifier electronics.
The inner shieldings and detectors were covered by a plastic
bag connected to the exhaust line of the dewar, serving
as a purge for the radioactive radon gas.

\begin{figure}[hbt]
\includegraphics[width=8cm]{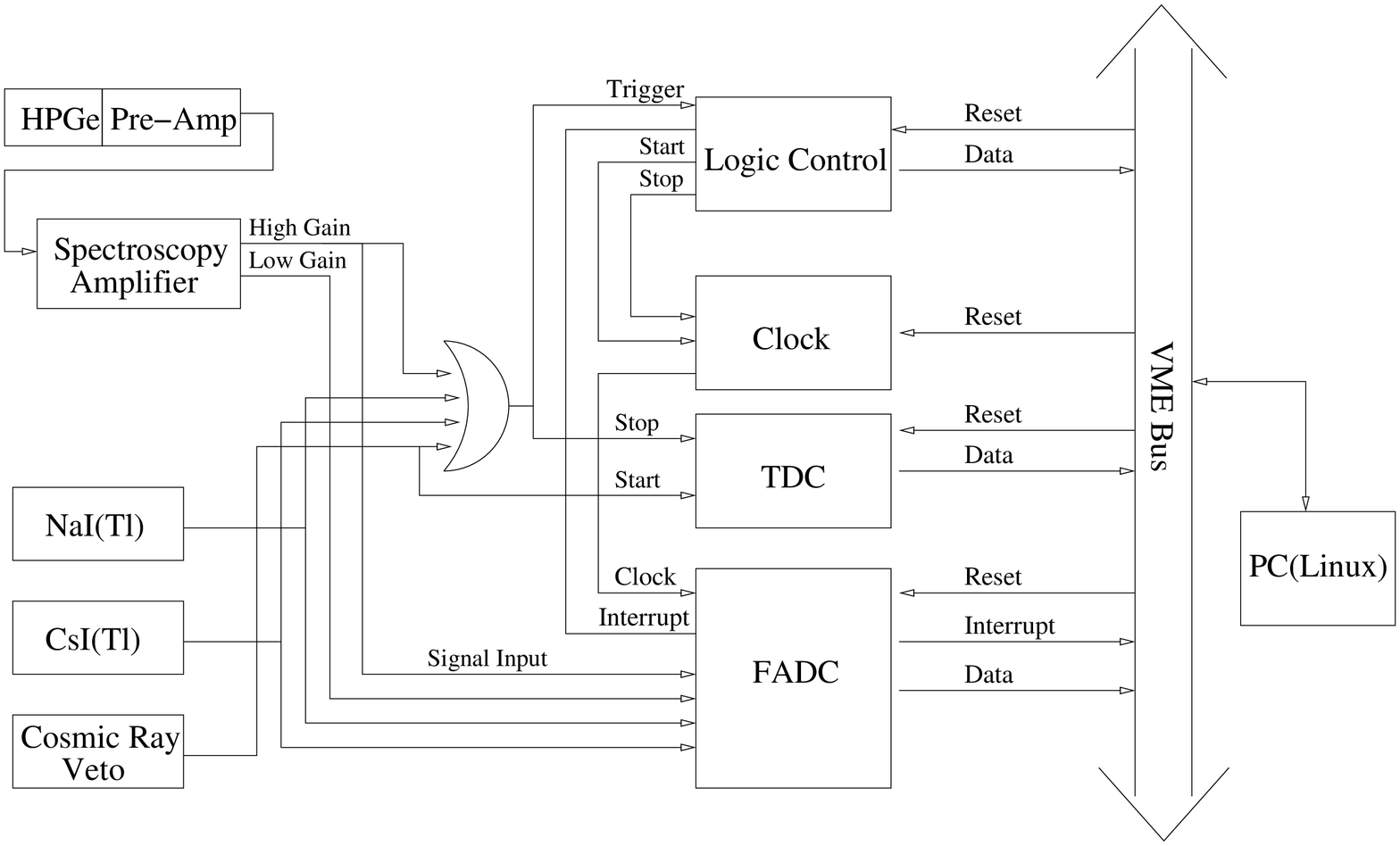}
\caption{
Schematic layout of the 
electronics and data acquisition
systems of the HPGe and the associated
ACV and CRV detectors. 
}
\label{daqblock}
\end{figure}

The electronics and data acquisition 
(DAQ) system~\cite{eledaq} of 
the HPGe detector assembly is illustrated
schematically in Figure~\ref{daqblock}.
The HPGe pre-amplifier signals were distributed to
two spectroscopy amplifiers
at the same 4~$\mu$s shaping time but with different
gain factors.
The amplifier signals were fed to a 
discriminator with a minimal threshold.
The discriminator output 
provided the on-line triggers, 
ensuring all the events
down to the electronics noise edge of 5~keV were recorded.
The amplifier output of the HPGe
and the PMT signals
from the ACV detectors  were recorded by 20~MHz Flash
Analog to Digital Convertor (FADC) modules
with 8-bit dynamic range
for a duration of 10~$\mu$s and 25~$\mu$s before and
after the trigger, respectively.
The discriminator output for the various
channels of all three systems (HPGe, ACV, CRV),
as well as the timing
output of the CRV PMTs, were
also recorded.
The redundancy of the readout channels enhanced
robustness and stability to the detector performance.
A random trigger (RT) was generated by an external
clock at a sampling rate of 0.1~Hz.
The RT events provided accurate
measurements of the pedestal levels,
the DAQ dead time as well as
the various efficiency factors.
The data were read out via the VME bus
through an VME-PCI interface~\cite{vme}
to a PC running on Linux operating system.
The data were saved on hard disks.
The DAQ system remained active for 2~ms after
a trigger to  record possible time-correlated signatures.
The typical data taking rate for the HPGe sub-system
was about 1~Hz. The DAQ dead time was about
4~ms and 2~ms per event for the HPGe 
and CsI(Tl) triggers, respectively.
The system live times 
are listed in Table~\ref{daqperiod}.
They varied among the three periods
because of the different trigger rates.

The KS Laboratory is connected with the
control center at Academia Sinica (AS)
through the telephone lines via MODEM.
Internet connections are prohibited due to security
reasons. During steady-state data taking, 
the experiment operated automatically without
the necessity of human presence.
The operating conditions
were constantly monitored via the MODEM.
Stability of key parameters 
like temperatures, liquid nitrogen levels,
trigger rates and DAQ live times
were checked.
The KS Nuclear Power Station can be reached by an hour's
drive from AS. The
laboratory was visited typically once a week,
when manual checks and calibrations were performed.
Hard disks were retrieved and brought back to AS
where data were archived and copied.
A total of $\sim$125~Gbytes of raw data were recorded 
in the three data taking periods by the HPGe system.
The data to be processed were installed on to
an external disk array storage system~\cite{raid} 
with 1.6~Tbytes
total storage capacity.
Relevant and high-level data on 
physical quantities were then
extracted and distributed among the users,
on which data analysis were performed.

\section{VI. Data Analysis and Understanding} 

Scatterings of $\nuebar$-e inside the Ge target
would manifest as ``lone-events'' uncorrelated
with other detector systems.
These events were extracted from the raw data
through selection criteria including
pulse shape analysis (PSA), anti-Compton (ACV) 
and cosmic-ray vetos (CRV).

\begin{figure}[hbt]
{\bf (a)}\\
\includegraphics[width=8cm]{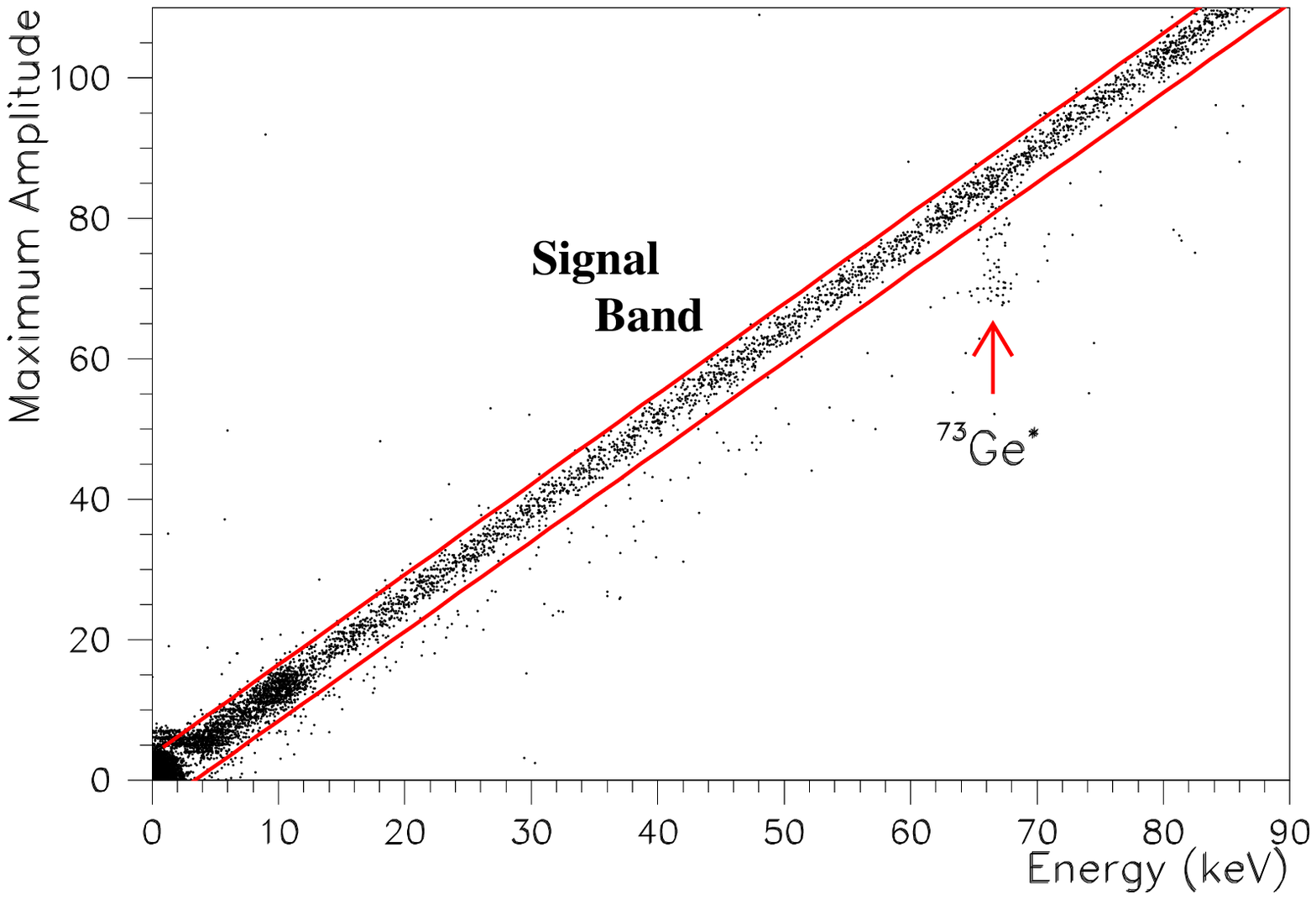}\\
{\bf (b)}\\
\includegraphics[width=8cm]{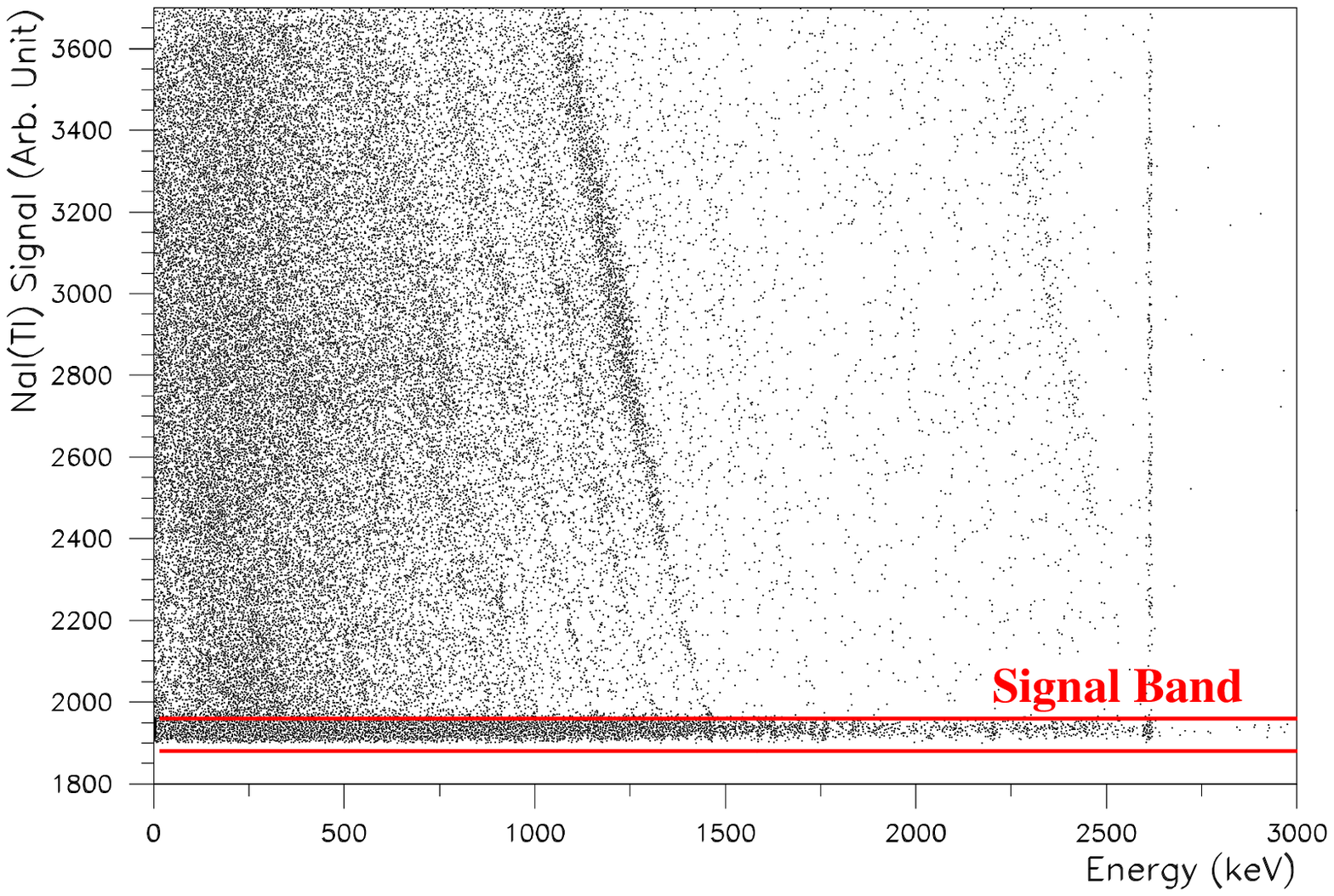}\\
{\bf (c)}\\
\includegraphics[width=8cm]{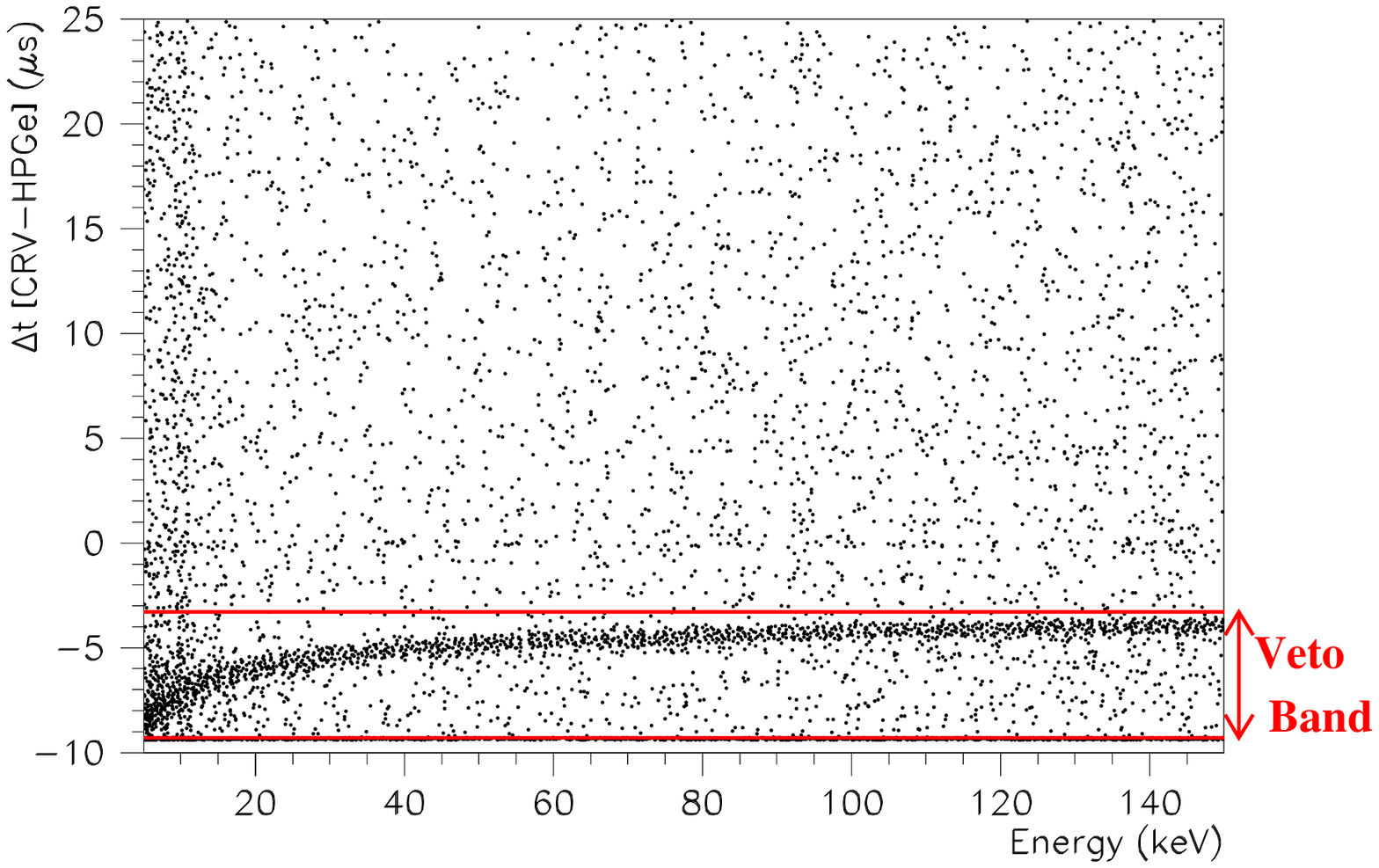}
\caption{
Selection procedures of the recorded data :
(a) pulse shape analysis, (b) anti-Compton selection,
and (c) cosmic-ray veto.
}
\label{selection}
\end{figure}

As displayed in the correlation plot between
pulse area and amplitude in Figure~\ref{selection}a, 
spurious background due to accidental
and delayed ``cascade'' events were suppressed
by PSA.
In particular, the structure at 66~keV was from
the decays of $^{73}$Ge$^*$ via the emissions of
two photons separated by a half-life of 4.6~$\mu$s.
The ACV and CRV cuts suppressed Compton scattering
and cosmic-ray induced events, respectively.
The scattered plot of HPGe and NaI(Tl) well-detector
events  is shown in Figure~\ref{selection}b.
A software threshold of about 5~keV was
adopted for the ACV detector to define the veto 
regions.
The various slanted bands were due to full energy 
depositions of $\gamma$-rays in the two detectors,
such as those from $^{40}$K and the $^{238}$U
and $^{232}$Th series.
The vertical band at 2614~keV originated from 
the coincidence of the $\gamma$-ray cascade
following the decays of $^{208}$Tl.
The timing correlations between CRV and HPGe events 
are shown in Figure~\ref{selection}c.
The PMT pulses from the CRV 
were fast (10~ns rise time) and therefore
arrived before the HPGe signals which  
were processed through amplifiers with
4~$\mu$s shaping time.
The time difference increased with lower energy  
because the trigger timing was defined by
a constant-threshold discriminator. 
The selected regions for the
lone-events are displayed
in the respective plots in Figure~\ref{selection}.

The background suppression factors 
as well as the signal survival efficiencies 
for the
three data taking periods are summarized in
Table~\ref{tabselect}.
The RT events were uncorrelated to the other
parts of the detector systems $-$ similar to
the neutrino-induced events.
The survival probabilities of the
RT events along the various stages of the
analysis procedures provided accurate measurements
of the DAQ and analysis efficiencies.
The DAQ live time
is the ratio between 
RT events actually recorded on disk
to the total numbers of RT signals 
generated by the clock.
The ACV and CRV selection efficiencies are
the fractions of the RT events 
which survived the cuts. 
Only loose cuts of 
$\pm 5 \sigma$ around the signal
band were used in the PSA cut as shown
in Figure~\ref{selection}a,
such that  
its efficiency is close to unity.

\begin{table*}[hbt]
\caption{\label{tabselect}
Summary of the event selection procedures
as well as their background
suppression and signal efficiency factors.
The survival probabilities are defined as
the ratio between the events selected
to those prior to these cuts but after previous
cuts were applied.
}
\begin{ruledtabular}
\begin{tabular}{lccc|ccc}
Event Selection &
\multicolumn{3}{c|}{Background Suppression} &
\multicolumn{3}{c}{Signal Efficiency}  \\ \hline
Period & I & II & III & I & II & III \\ \hline
Raw Data  & 1.0 & 1.0 & 1.0 & 1.0 & 1.0 & 1.0 \\
Pulse Shape Analysis (PSA)  & $>$0.99  & $>$0.99
& $>$0.99  & $>$0.99  & $>$0.99  & $>$0.99 \\
Anti-Compton Veto (ACV)  & 0.054 & 0.051 & 0.058 & 0.99 & 0.99 & 0.99  \\
Cosmic-Ray Veto (CRV) & 0.92 & 0.85 & 0.80 &  0.95 & 0.94 & 0.93 \\ \hline
Combined Efficiency & 0.050 & 0.043 & 0.046 & 0.95 & 0.93 & 0.92
\end{tabular}
\end{ruledtabular}
\end{table*}

\begin{figure}[hbt]
\includegraphics[width=8cm]{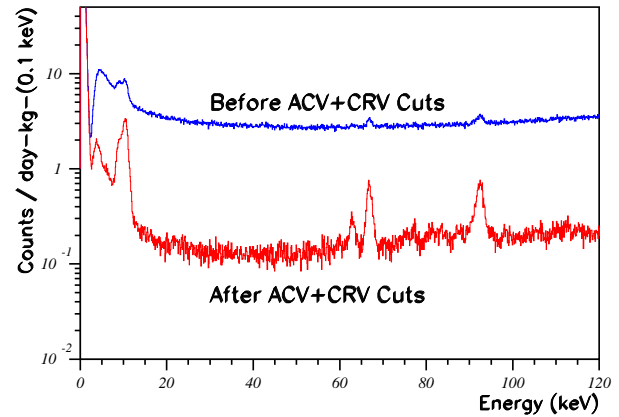}
\caption{
Measured
spectra before and after the ACV+CRV cuts
for the Reactor ON data in Period-III.
}
\label{rawspect}
\end{figure}

\begin{figure}[hbt]
{\bf (a)}\\
\includegraphics[width=8cm]{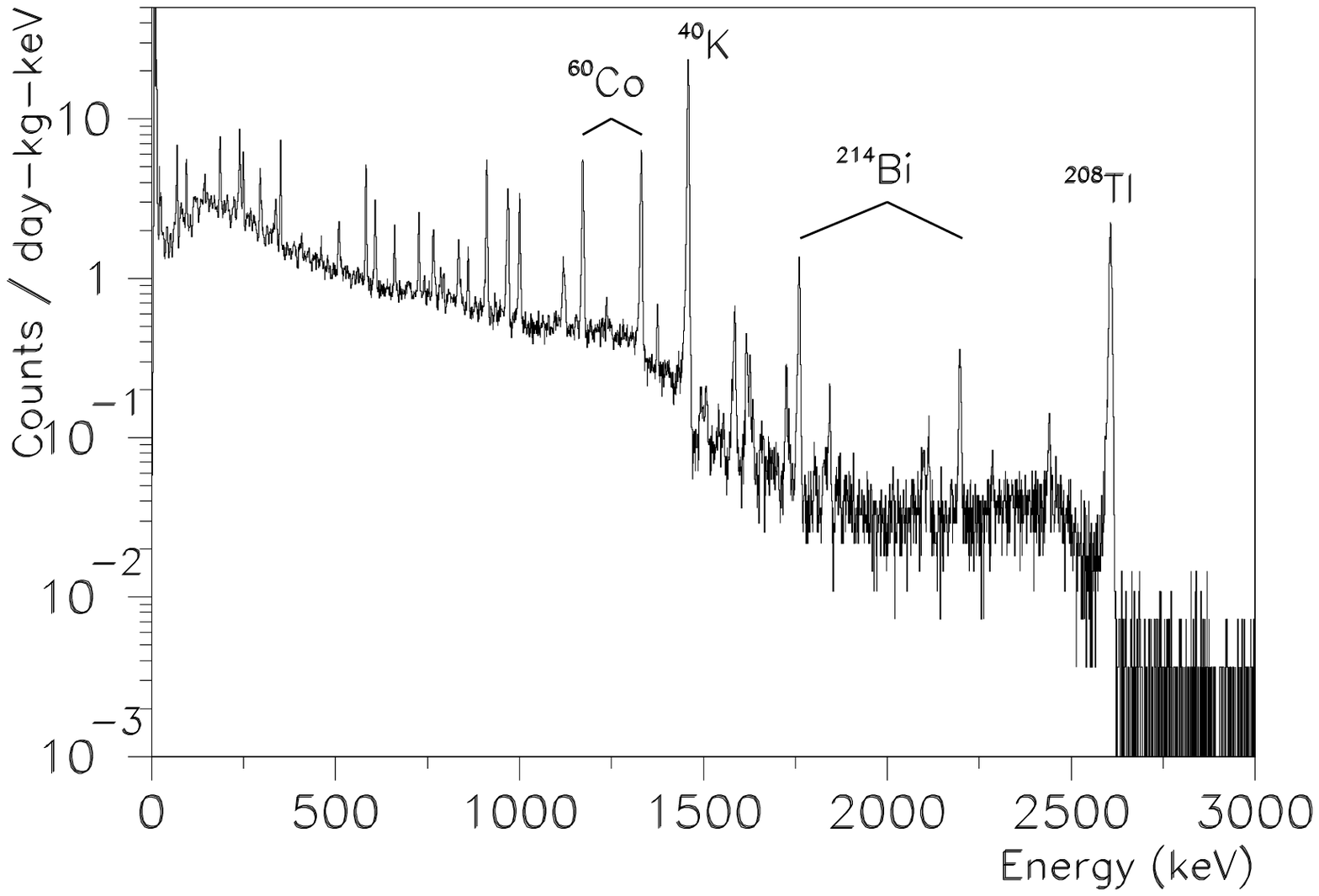}\\
{\bf (b)}\\
\includegraphics[width=8cm]{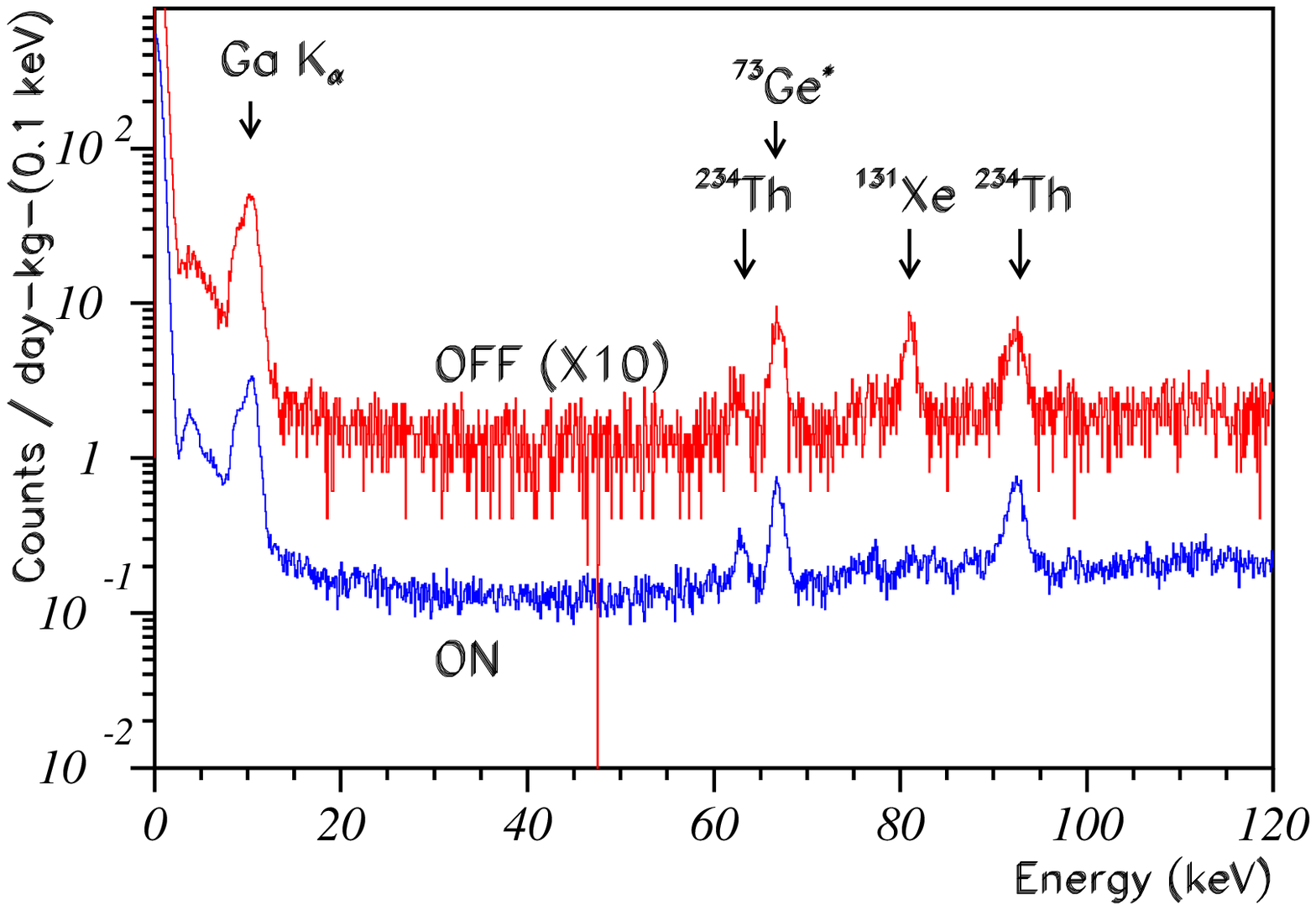}\\
\caption{
Measured spectra in Period-III after all the
selection criteria were applied for:
(a) the full energy range
and (b) the range below 120~keV relevant
to magnetic moment studies.
Key $\gamma$-lines were identified.
Both Reactor ON and OFF spectra are
separately displayed in (b).
}
\label{finalspectrum}
\end{figure}

As illustrations, the 
measured spectra in Period-III 
before and after the ACV+CRV cuts for the
Reactor ON data are depicted in Figure~\ref{rawspect}.
The spectra after the event selection
criteria, also for Period-III,
are displayed in Figures~\ref{finalspectrum}a 
and \ref{finalspectrum}b  for the high and low
energy ranges, respectively.
The main $\gamma$-lines were identified and tabulated
in Table~\ref{tabgamma}.
Most of the $\gamma$-activities
were due to natural background from 
the $^{238}$U and $^{232}$Th series,
as well as from $^{40}$K and $^{235}$U.
The existence of activities from $^{235}$U
is particularly instructive. The low-energy 
lines implies that the source should be close
to the target, most probably inside the OFHC
cryostat, where $^{238}$U and $^{232}$Th should
also be present in larger amount. 
These activities were likely to originate from the 
front-end pre-amplifier components located
in the vicinity of the HPGe target. These were
expected to be one of the
important background sources. 

\begin{table}[hbt]
\caption{\label{tabgamma}
Summary of $\gamma$-lines intensity measured in Period-III. }
\begin{ruledtabular}
\begin{tabular}{rcccr}
Energy & Isotopes & Source / & $\halflife$ & Intensity \\
(keV) & & Decay Series & & ($\rm{kg ^{-1} day^{-1}}$) \\ \hline
   66.7 &  $^{73m}$Ge  &    cosmic 
    &          0.5 s         &  15.4$\pm$0.4 \\
   92.6 &  $^{234}$Th  &  $^{238}$U  &         24.1 d         &  11.9$\pm$0.5 \\
  143.8 &  $^{235}$U   &  $^{235}$U  &  7.0$\times 10^{8}$ y  &   5.1$\pm$0.8 \\
  185.7 &  $^{235}$U   &  $^{235}$U  &  7.0$\times 10^{8}$ y  
&  \multirow{2}*{\} 17.2$\pm$0.4} \\
  186.2 &  $^{226}$Ra  &  $^{238}$U  &         1600 y         &  \\ 
  238.6 &  $^{212}$Pb  &  $^{232}$Th &         10.6 h         &  18.8$\pm$0.5 \\
  249.8 &  \multicolumn{3}{c}{unidentified}      &  11.6$\pm$0.5 \\
  295.2 &  $^{214}$Pb  &  $^{238}$U  &         26.8 m         &   6.3$\pm$0.3 \\
  338.3 &  $^{228}$Ac  &  $^{232}$Th &          6.2 h         &   3.7$\pm$0.5 \\
  351.9 &  $^{214}$Pb  &  $^{238}$U  &         26.8 m         &  17.1$\pm$0.4 \\
  463.0 &  $^{228}$Ac  &  $^{232}$Th &          6.2 h         &   1.6$\pm$0.3 \\
  583.2 &  $^{208}$Tl  &  $^{232}$Th &          3.1 m         &  14.4$\pm$0.3 \\
  609.3 &  $^{214}$Bi  &  $^{238}$U  &         19.9 m         &   8.1$\pm$0.2 \\
  661.7 &  $^{137}$Cs  &   CsI(Tl)     
&         30.1 y         &   4.6$\pm$0.2 \\
  727.3 &  $^{212}$Bi  &  $^{232}$Th &         66.6 m         &   6.4$\pm$0.2 \\
  766.4 &  $^{234m}$Pa &  $^{238}$U  &          1.2 m         &   5.0$\pm$0.3 \\
  785.4 &  $^{212}$Bi  &  $^{232}$Th &         66.6 m         &  
\multirow{2}*{\} 1.7$\pm$0.4} \\
  786.0 &  $^{214}$Pb  &  $^{238}$U  &         26.8 m         &  \\
  795.0 &  $^{228}$Ac  &  $^{232}$Th &          6.2 h         &   2.7$\pm$0.8 \\
  834.8 &  $^{54}$Mn   &    reactor   
&        312.3 d         &   3.6$\pm$0.3 \\
  860.6 &  $^{208}$Tl  &  $^{232}$Th &          3.1 m         &   3.5$\pm$0.3 \\
  911.2 &  $^{228}$Ac  &  $^{232}$Th &          6.2 h         &  19.1$\pm$0.3 \\
  964.8 &  $^{228}$Ac  &  $^{232}$Th &          6.2 h         &  
\multirow{2}*{\} 14.4$\pm$0.3} \\
  969.0 &  $^{228}$Ac  &  $^{232}$Th &          6.2 h         &  \\
 1001.0 &  $^{234m}$Pa &  $^{238}$U  &          1.2 m         &  11.4$\pm$0.3 \\
 1120.3 &  $^{214}$Bi  &  $^{238}$U  &         19.9 m         &   6.7$\pm$0.5 \\
 1173.2 &  $^{60}$Co   &    reactor     
&          5.3 y         &  26.0$\pm$0.3 \\
 1238.1 &  $^{214}$Bi  &  $^{238}$U  &         19.9 m         &   1.2$\pm$0.2 \\
 1332.5 &  $^{60}$Co   &     reactor     
&          5.3 y         &  27.0$\pm$0.3 \\
 1377.7 &  $^{214}$Bi  &  $^{238}$U  &         19.9 m         &   1.9$\pm$0.3 \\
 1460.8 &  $^{40}$K    &   natural   
&  1.3$\times 10^{8}$ y  & 106.4$\pm$1.0 \\ 
 1509.2 &  $^{214}$Bi  &  $^{238}$U  &         19.9 m         &   0.6$\pm$0.1 \\
 1588.2 &  $^{228}$Ac  &  $^{232}$Th &          6.2 h         &   2.5$\pm$0.1 \\
 1620.5 &  $^{212}$Bi  &  $^{232}$Th &         66.6 m         &   1.6$\pm$0.1 \\
 1630.6 &  $^{228}$Ac  &  $^{232}$Th &          6.2 h         &   0.6$\pm$0.1 \\
 1729.6 &  $^{214}$Bi  &  $^{238}$U  &         19.9 m         &   1.1$\pm$0.1 \\
 1764.5 &  $^{214}$Bi  &  $^{238}$U  &         19.9 m         &   5.9$\pm$0.9 \\
 1847.4 &  $^{214}$Bi  &  $^{238}$U  &         19.9 m         &   0.7$\pm$0.3 \\
 2118.6 &  $^{214}$Bi  &  $^{238}$U  &         19.9 m         &   0.2$\pm$0.1 \\
 2204.2 &  $^{214}$Bi  &  $^{238}$U  &         19.9 m         &   2.3$\pm$0.4 \\
 2447.9 &  $^{214}$Bi  &  $^{238}$U  &         19.9 m         &   0.5$\pm$0.1 \\
 2614.5 &  $^{208}$Tl  &  $^{232}$Th &          3.1 m         &  14.5$\pm$0.2 \\
\end{tabular}
\end{ruledtabular}
\end{table} 

Besides natural radioactivity,
there were evidence of
$^{60}$Co and $^{54}$Mn, both being known contaminations
within the reactor building.
Decay profiles were observed for $^{54}$Mn consistent
with its nominal decay life-time.
The peak at 662~keV was due to
$^{137}$Cs activity in
the CsI(Tl) ACV detector located near
the HPGe.
There was an unidentified line at 249.8~keV.
The intensity was uniform to $<$4\%
in both ON and OFF periods but
the short-duration rates
fluctuated more than the
statistical uncertainties would allow.
No interpretations consistent with
the other spectral features could be found.

The various $\gamma$-lines
also provided in situ energy calibration as well as
independent measurements and
consistency checks to
the stability and efficiency factors
complementary to the RT events.
Details on background and detector 
stabilities will be presented in
Section~VII in connection
with the discussions on
the systematic uncertainties.

The low energy spectrum of
Figure~\ref{finalspectrum}b is relevant to the
studies of neutrino magnetic moments.
A detector threshold of 5~keV
and a background of
$\sim$1 keV$^{-1}$kg$^{-1}$day$^{-1}$
above 12~keV were achieved.
The background level is comparable to the 
typical range
in underground Cold Dark Matter experiments.
This is a notable achievement for 
an experiment at shallow-depth $-$ {\it and}
in the vicinity of a power reactor core.
Several $\gamma$-lines can be identified:
(a) Ga X-rays at 10.37~keV and $^{73}$Ge$^*$ at 66.7~keV
from internal cosmic-induced activities,
(b) $^{234}$Th at 63.3~keV and 92.6~keV 
from the $\u238$ series due to residual ambient
radioactivity close to the target, 
and (c) a line at 80.9~keV in the OFF spectrum,
to be examined in a subsequent paragraph.
The copper cryostat as well as the inactive
surface electrode of the HPGe provided total 
suppression to external low energy photons.
Consequently, the measured spectra 
did not exhibit any structures above the
the Ga X-rays end point at 12~keV up to
about 60~keV.

\begin{figure}[hbt]
{\bf (a)}\\
\includegraphics[width=8cm]{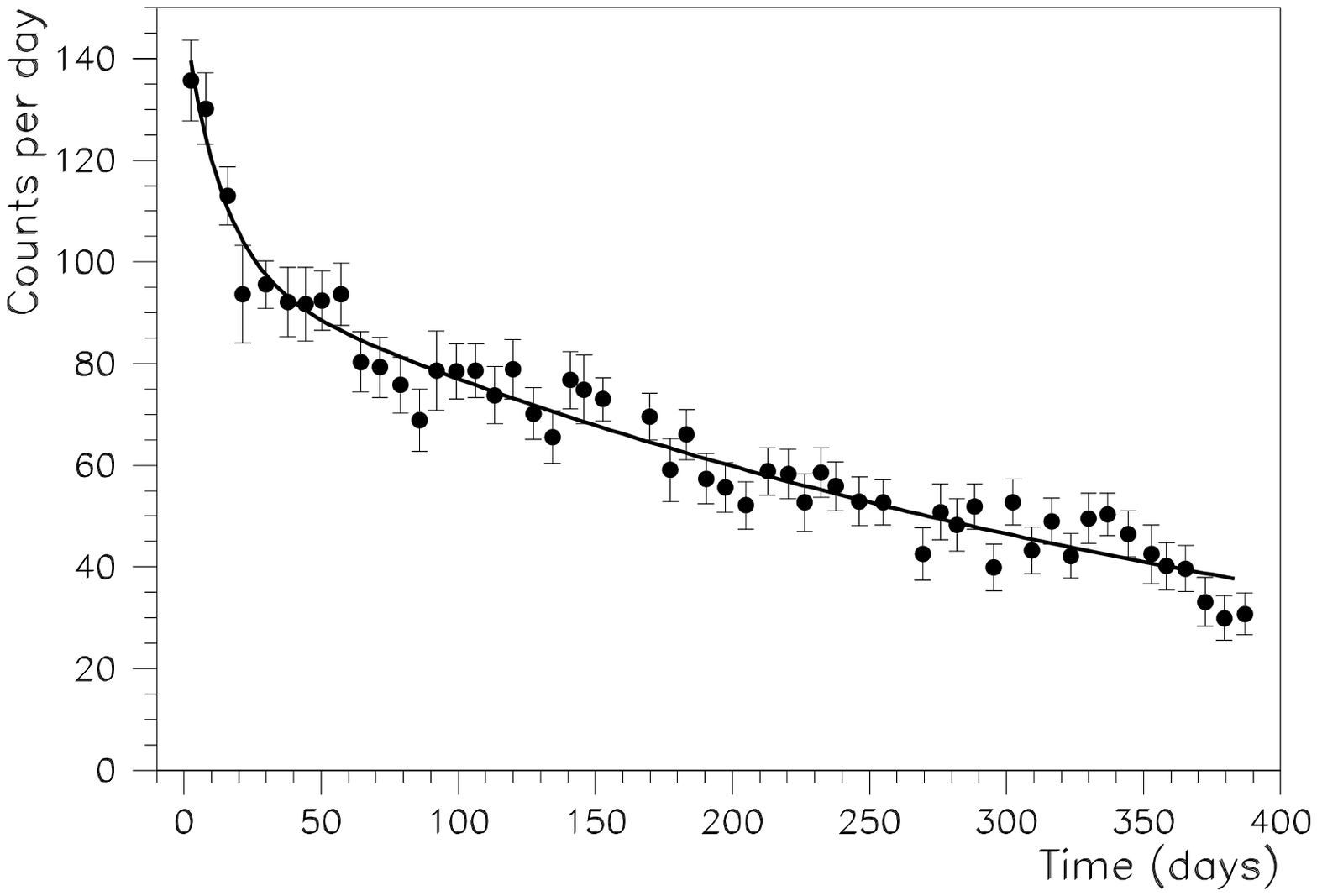}\\
{\bf (b)}\\
\includegraphics[width=8cm]{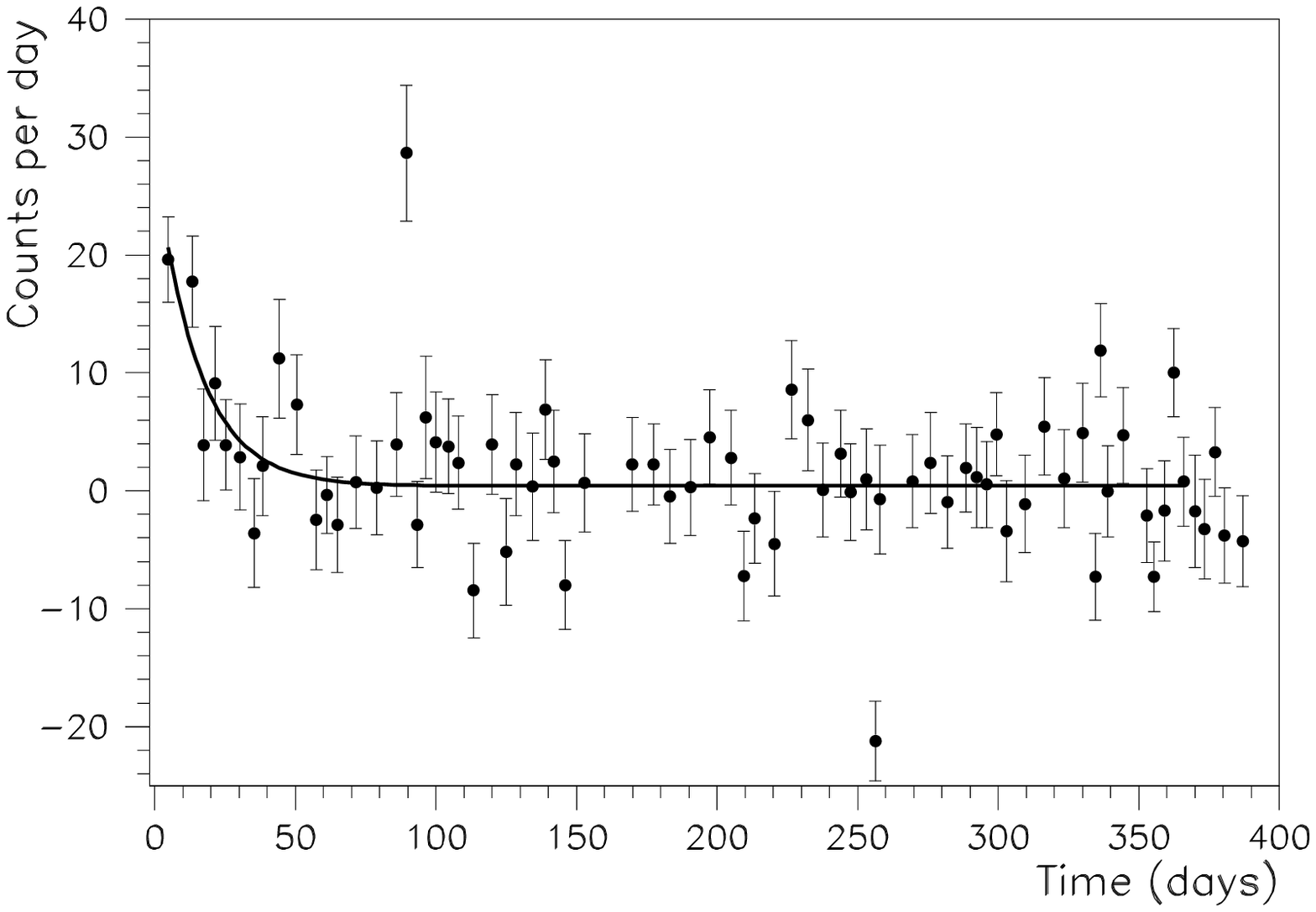}\\
\caption{
The time-variation plots
in Period-III data taking
for $\gamma$-lines
due to
(a) Ga X-rays
at 10.37~keV and (b) the 80.9~keV peak interpreted as
$\gamma$'s  from $^{133}$Xe.
}
\label{stability}
\end{figure}

Time variations of two features were observed
in the low energy spectrum displayed in 
Figure~\ref{finalspectrum}b.
The intensity of the
10.37~keV Ga X-ray peak 
decayed with time, as
illustrated in Figure~\ref{stability}a
in the case of data from Period-III.
The best-fit half-lives of 
275$\pm$14~days  and
9.6$\pm$3.0~days 
to a double-exponential function
agree well with the decays of 
$^{68}$Ge and $^{71}$Ge both of
which were neutron-activated:
\begin{equation}
\rm{
^{70}Ge (n,3n) ^{68}Ge   :  ~
^{68}Ge + e^-  \rightarrow  ^{68}Ga  ~ 
( \halflife = 270.8 ~  days ) 
}
\end{equation}
and
\begin{equation}
\rm{
^{70}Ge (n, \gamma ) ^{71}Ge   :  ~ 
^{71}Ge  + e^-  \rightarrow  ^{71}Ga  ~ 
( \halflife = 11.4 ~ days ) ~.
}
\end{equation}
The observed exponential decays
indicated that there was less neutron
activation on the Ge-target inside the
shielding structures at KS compared
to the AS laboratory where the detector hardware was
prepared and tested unshielded 
prior to installation.
Coupled with these processes were the
$\beta ^+$ decays of $^{68}$Ga (Q=2.9~MeV
with 86\% branching ratio). They were
associated with two 511~keV photons  emitted
in coincidence, such that the ACV suppression
was large.


Another interesting feature is the
presence of a line at 80.9~keV observed
{\it only once}
in the beginning of the Period-III data taking,
which was a Reactor OFF duration.
The time evolution is displayed in Figure~\ref{stability}b.
The best-fit half-life of 10.6$\pm$3.3~days
is consistent
with the interpretation of
$\gamma$-emissions following
$\beta$-decays of $^{133}$Xe ($\halflife$=5.24~days).
This isotope is a fission fragment with
large cumulative yield ($\sim$7\%).
It exists in gaseous form and has long enough 
half-life to leave the reactor core. 
Air pockets contaminated with $^{133}$Xe
might have been trapped in the vicinity of the
detector during installation.
Photons at this low energy  were fully absorbed
absorbed in the HPGe, such the time variations
were confined to the peak region. 

The comparisons of the key features in the measured
spectra from the
three periods are given in Table~\ref{tabcompare}.
There were no hardware intervention to the shielding
structures and the detectors during data taking within
one period to ensure stability of the ambient conditions
and of the detector operations.
In between the data taking periods,
there were maintenance, hardware
improvement as well as detector installation efforts.
The shielding door was
opened and the radon purge system was 
temporarily disconnected.
The passive shielding configurations in
the inner target volume were re-assembled.
All these operations were performed during
Reactor ON situations to avoid the
possibilities of contaminations 
in reactor outage.
There were no incidents of sudden
surge of any background during the ON-to-OFF transitions
from both the HPGe detector as well as the CsI(Tl) array 
in all the three periods.

It can be seen from Table~\ref{tabcompare}
that the $^{40}$K activity
and the before-cut background level at 30~keV
associated with the
experimental hardware
were maintained stable during
the different data taking periods.
The HPGe system had a 
hardware failure 
at the end of Period-II data taking,
which required the detector to be shipped
back and forth by air flights 
to the manufacturer for repairs.
Consequently, there was a sharp rise in 
the cosmic-induced
Ga X-ray intensity between Periods II and III.
The reduction 
in the $^{208}$Tl background between Periods II
and III might be due to the replacement
of pre-amplifier electronic
components necessary for the repairs.
The background due to the $^{60}$Co and $^{54}$Mn
contaminations at the reactor 
were much suppressed by the shieldings
and experimental precautions, such that
the detector-associated $^{40}$K
became the dominant $\gamma$-activity. 

\begin{table*}[hbt]
\caption{\label{tabcompare}
Summary of the major
spectral measurements in the three data taking
periods. 
The intensities of the Ga X-rays 
are those at the start of the data taking.
The others are time-averaged over an entire
period. 
}
\begin{ruledtabular}
\begin{tabular}{lccccc|cc}
\underline{Period} & 
\multicolumn{5}{c|}{\underline{Event Rates ($\rm{kg ^{-1} day ^{-1}}$)}} &
\multicolumn{2}{c}{\underline{30~keV Band ($\cpd$)}}\\
& Ga X-Rays & $^{40}$K  & $^{208}$Tl & $^{60}$Co & $^{54}$Mn ~ & 
Before-Cut  & After-Cut  \\ \hline
I & 134$\pm$6 & 107.2$\pm$0.7 & 29.2$\pm$0.3 & 9.8$\pm$1.0 & 4.2$\pm$0.7 ~
& 31.6$\pm$0.1 & 1.58$\pm$0.02 \\
II & 45$\pm$3 & 106.4$\pm$0.9 & 25.6$\pm$0.4 & 14.5$\pm$1.7 & 
3.1$\pm$0.4 ~  &
31.2$\pm$0.1 & 1.34$\pm$0.02 \\
III & 128$\pm$7 & 106.4$\pm$1.0 & 14.5$\pm$0.2 & 26.5$\pm$0.2 & 3.6$\pm$0.3 ~ & 
30.9$\pm$0.1 & 1.42$\pm$0.01    \\ \hline
RMS/Mean & 0.40 & 4e-3 & 0.27 & 0.42 & 0.12 & 9e-3  &  0.07
\end{tabular}
\end{ruledtabular}
\end{table*}

\section{VII. Neutrino Magnetic Moments and Radiative Decay Searches}
\label{sect::mm}

Only the low energy spectra like the one
shown in Figure~\ref{finalspectrum}b
were used in the studies of $\munu$.
Since data taking conditions were not identical among the
three periods, the data were analyzed 
and Reactor ON/OFF spectra 
were compared independently within
each period.
Consequently, the variations in the key
background features shown in Table~\ref{tabcompare}
do not affect the evaluations of $\munu$.
The analysis procedures were extended from those used
previously~\cite{prl03} through a global
treatment which 
incorporated additional constraints
and measurements.

The PSA cut identified  events with spurious and convoluted
pulse shapes, some of which were due to electronic noise. 
These were not stable with time in general.
The ACV and CRV cuts, however, rejected background
due to physical processes such that the suppression
factors should be stable during data taking provided
that the
hardware were properly controlled and cross-monitored.
We denote 
the spectra surviving the PSA cut 
by $\phi_{-}^i$ and $\phi_{+}^i$ for the
Reactor OFF and ON data, respectively,
while those after further ACV+CRV cuts similarly by
$\phi_{-}^f$ and $\phi_{+}^f$.
The before and after-cut Reactor ON spectra 
$\phi_{+}^i$ and $\phi_{+}^f$ in Period-III are
displayed in Figure~\ref{rawspect}.
The neutrino-induced contributions to
the electron recoil spectra
can be  described by
\begin{equation}
\nu =
 \Phi _e (SM) + \ke10 ^2 \cdot \Phi _e ( \munu = 10^{-10} ~ \mub ) ~ .
\end{equation}
The $\ke10 ^2$ dependence on the $\munu$-signals
follows from Eq.~\ref{eq::mm}.

There are 
two constraints relating
the four measured spectra within
one period $-$ 
\begin{description}
\item{(I)}
the excesses of $\phi_{+}^f$ over $\phi_{-}^f$, if any, 
are neutrino-induced:
\begin{equation}
\label{eq1}
\phi_{+}^f  ~ =  ~ \epsilon \cdot \nu + \phi_{-}^f
\end{equation}
where $\epsilon$ is 
the selection efficiency of the ACV+CRV cuts
as given in Table~\ref{tabselect}, and 
\item{(II)}
the suppression factors 
for the ACV+CRV cuts
should be constant between
the Reactor-ON and OFF data taking:
\begin{equation}
\label{eq2}
\frac{ \phi_{+}^f - \epsilon \cdot \nu }{ \phi_{+}^i - \nu } ~ = ~
\frac{ \phi_{-}^f }{ \phi_{-}^i }  ~.
\end{equation}
\end{description}

Global minimum-$\chi ^2$ analyses were
performed to the data from the three periods 
independently. 
Though the DAQ threshold was 5~keV, 
the analysis threshold of 12~keV was chosen to 
avoid complications of the atomic effects~\cite{atomic}
as well as the time-varying background from Ga X-rays.
The background profile 
was continuous and stable from
this energy threshold up to 61~keV as expected,
and could be described by a 
fit of a polynomial function
to $\phi_{-}^f$.
The best-fit values of $\ke10 ^2$ as well as 
the $\chi ^2$/dof 
for the three periods are listed in Table~\ref{fitresults}.
It can be seen that Eqs.~\ref{eq1}
and \ref{eq2} provide excellent descriptions to
the data.

\begin{table}[hbt]
\caption{\label{fitresults}
Summary of the best-fit results
for the three periods
and the $\chi ^2$/dof
indicating the goodness-of-fit.
}
\begin{ruledtabular}
\begin{tabular}{lcc}
\underline{Period} & 
\multicolumn{2}{c}{\underline{Best-Fit}}\\
&  $\ke10 ^2$ &  $\chi ^2$/dof \\ \hline
I &   -0.52$\pm$1.05 &  90/97 \\
II & 0.06$\pm$1.21 & 108/97  \\
III & -0.84$\pm$0.87 & 84/97 \\ \hline
Combined &  -0.53$\pm$0.59 & $-$
\end{tabular}
\end{ruledtabular}
\end{table}

The data taking durations of the three
periods lasted from eight to twelve months.
Demonstrations of the background and detector
stabilities are therefore crucial.
Since each period was treated independently
with its own Reactor ON and OFF data compared,
the variations among the different periods
shown in Table~\ref{tabcompare}
did not lead to systematic effects on the analysis.
Nevertheless, the stability 
(an RMS spread of 0.9\%)  of the
before-cut levels at 30~keV
among the different periods 
suggested that the overall background was in
good control throughout the experiment.

Two complementary studies were performed on
the systematic effects of the experiment,
and they are discussed in details in the
following paragraphs.

\begin{table}[hbt]
\caption{\label{stabmonitor}
The measured stability levels of various
key spectral features, based on P-III data:
R$_{\pm}$ is
the ``residual/statistical error'' ratio
for the ON/OFF data,
while $\delta _{tot}$ is the error/mean ratio
for the entire DAQ period.
}
\begin{ruledtabular}
\begin{tabular}{lccr}
Monitors & R$_{\pm}$ & $\delta _{tot}$
& $\chi ^2$/dof \\ \hline
$^{40}$K 1462~keV Band & -0.42 & 3.5e-3 & 55/53 \\
$^{208}$Tl 2614~keV Band  & 0.18 & 1.3e-2 & 50/53 \\
Band at 30 keV  & 0.005 & 8.2e-3 & 55/53 \\
Band at 400 keV  & 1.3 & 4.1e-3 & 79/73\\
Suppression factor  &  &  \\
$~~$ (ACV \& CRV) at 30 keV & -0.21  & 9.6e-3 & 72/73 \\
\end{tabular}
\end{ruledtabular}
\end{table}

{\noindent
\bf (1) Stability Measurements:}

Several important spectral features  
were continuously measured to monitor
and demonstrate the stabilities of the
detector performance and background conditions.
These stabilities are summarized in Table~\ref{stabmonitor}
for the 350~days of Period-III data.
The monitored features included
the (ACV+CRV) suppression factors, 
as well as the event rates of 
the $^{40}$K and $^{208}$Tl bands,
together with the two sampling bands
at 30~keV and 400~keV.

To quantify the stabilities, the time-evolution
over the entire period was fit to a constant. 
The stability parameter $\delta _{tot}$ corresponds to
the ratio of error/mean of the fits.
The excellent values on $\chi ^2$/dof indicate
that the stability hypothesis is valid and
that the fluctuations of individual data points
were statistical in nature.  Accordingly,
$\delta _{tot}$ can be taken as the 
stability levels of the various features
being monitored.
The parameter R$_{\pm}$ denotes
the ratio of residuals of Reactor ON$-$OFF data
to their respective 1$\sigma$ statistical accuracies. 
The measured results of  $| R _{\pm} | \sim 1$ 
indicate that the experimental conditions were stable
between the Reactor ON/OFF data set, such that
their differences were consistent with 
statistical fluctuations, rather than due to
systematic variations in hardware or background.

\begin{table}[hbt]
\caption{\label{bounds}
Identified candidate channels which may 
contribute to Reactor ON/OFF instabilities.
Their impact on Reactor ON/OFF comparisons
at the 30~keV signal region
($\delta _{\pm}$) were evaluated
from P-III data and simulations.
The three categories are listed separately:
(a) event rate changes,
(b) background variations, and
(c) instabilities in detector performance and
analysis efficiencies.
The intrinsic stabilities of the background
channels in Table (b) are given by 
$\delta$(bkg) while $\eta$(MC) 
denotes the simulated background level
at the signal region with keV$^{-1}$ bin-width,
normalized to one observed event at the peak.
Entries in Table (c) include 
the stabilities of the
detector systems $\delta$(Detectors) and
veto efficiencies $\delta$(Veto), while
the background fraction for each mode is given
by $f_B$.
The overall uncertainty limit combined in
quadratures is given in (d). 
}
{\bf (a) Event Rate Changes}\\
\begin{ruledtabular}
\begin{tabular}{lr}
Channels  &
ON/OFF Stability ($\delta _{\pm}$)  \\ \hline
DAQ live time  & $<$5e-4 \\
HPGe target mass   & $<$1e-4 \\ 
\end{tabular}
\end{ruledtabular}
\\[2ex]
{\bf (b) Background Variations}\\
\begin{ruledtabular}
\begin{tabular}{lccr}
Background  &
\multicolumn{3}{c}{\underline{ON/OFF Stability Levels}} \\
~~Source &  $\delta$(bkg) & $\eta$(MC)  &  $\delta _{\pm}$  \\ \hline
Radon diffusion &  $<$0.05  & $-$ & $<$5e-4\\
$^{54}$Mn  &  $<$0.3 &  2e-3 & $<$2e-3 \\
$^{60}$Co  &   $<$0.06 & 2e-3 & $<$4e-3 \\
$^{68}$Ga  &  $<$0.4 & 3e-8 & $<$1e-6 \\
$^{133}$Xe (only Period-III)   & 1 &  4e-4 & $<$3e-3 \\
Line at 249~keV  & $<$0.04 &  1e-3 & $<$3e-4 \\ 
\end{tabular}
\end{ruledtabular}
\\[2ex]
{\bf (c) Detector Performance and Data Analysis}\\
\begin{ruledtabular}
\begin{tabular}{lcccr}
Detector Systems  &  $f_B$   & 
\multicolumn{3}{c}{\underline{ON/OFF Stability Levels}} \\
~ \& Cuts &
&   $\delta$(Detector) &  $\delta$(Veto) &  $\delta _{\pm}$ \\ \hline
Both CRV \& ACV  & 0.64 & $<$5e-3 & $<$1e-6 &  $<$1e-5 \\
ACV only & 0.30  & $<$0.05 &  $<$7e-4 & $<$4e-3 \\
CRV only  & 0.01 & $<$0.1 & $<$0.01 & $<$2e-3 \\
Surviving CRV+ACV & 0.05 &  $<$0.15 &  $-$ & $<$5e-4 \\
\end{tabular}
\end{ruledtabular}
\\[2ex]
\begin{ruledtabular}
{\bf (d) Combined Limit}\\
\begin{tabular}{lr}
Combined (a)+(b)+(c) & $\delta _{\pm} <$ 7e-3 \\
\end{tabular}
\end{ruledtabular}
\end{table}

{\noindent
\bf (2) Bounds on Possible Instabilities:}

The various channels which may
contribute to possible instabilities
to the ON/OFF data at
the low energy signal region
were identified and their effects
were derived from actual data and 
complete detector
simulations. The results are summarized in
Table~\ref{bounds}a-c.
Their effects
to the ON/OFF comparisons at 30~keV, 
relative to the overall after-cut background 
$\phi \rm{ (cut)} \sim 1.5 ~ \cpd$,
are denoted by $\delta _{\pm}$.

The bounds 
on the ON/OFF stabilities 
due to event rate variations
are shown
in Table~\ref{bounds}a.
The DAQ live times were accurately
measured to $\delta _{\pm} <$5e-4
through the ratios of generated to
recorded RT events. 
Within each period, the HPGe detector was
continuously maintained at liquid nitrogen
temperature to minimize the diffusion of
the inactive layer which is about 1\% of the
active detector mass. A change of thickness
of $<$1\% implies a stability level of 
$\delta _{\pm} <$1e-4.

The various background sources which may produce
ON/OFF instabilities are listed in Table~\ref{bounds}b,
where their fractional variations
are denoted by $\delta$(bkg).
The evaporation rate of the liquid nitrogen
in the HPGe dewar was measured to be stable
to $<$1\%.
There were no observable changes 
[$\delta$(bkg)$<$5\%] in  
the key spectral features in a test data set
taken when this radon purge system was disconnected. 
Therefore, residual variations on the background
due to fluctuations 
in radon diffusion are expected to be
at the level of  $\delta _{\pm} <$5e-4.

There were five specific lines 
($^{54}$Mn, $^{60}$Co, $^{68}$Ga, $^{133}$Xe only for Period-III,
and the 249~keV line)
where variations with time were observed
or should be examined.
Their effects on the signal region 
were derived from 
full simulation studies using 
the measured peak intensities $\phi$(peak)
in Table~\ref{tabgamma}
and $\delta$(bkg) between the ON/OFF periods 
as input and normalizations.
The ON/OFF stabilities 
at the signal region are
the $\delta _{\pm}$ entries in
Table~\ref{bounds}b, derived from:
\begin{equation}
\delta _{\pm} \sim
[ ~ \phi {\rm (peak)} \cdot \delta \rm{ (bkg) } 
\cdot \eta {\rm (MC)} ~ ] ~ / ~ \phi \rm{ (cut) }
\end{equation}
where $ \eta {\rm (MC)}$ is 
the simulated background level 
at the low-energy signal region
with keV$^{-1}$ bin-width,
normalized to one observed event at the peak.
From physical expectations,
the background sources were taken 
to be uniformly distributed among the space
in the vicinity of the HPGe detector, 
and their emissions
were isotropic. 
The HPGe target was hermetically surrounded by 
the ACV and the detector response was mostly uniform
in all directions. 
Consequently, the effects due to
the exact source locations were small and 
the background levels only varied by $\sim$10\% (RMS)
if localized point sources were used instead.

The tight limits can be understood qualitatively
as follows. The weak intensity of the $^{54}$Mn line
and the long decay lifetime of the $^{60}$Co lines 
suppressed the ON/OFF variations 
to $\delta _{\pm} <$2e-3
and $\delta _{\pm} <$4e-3, respectively.
The decays of $^{68}$Ga are mostly via the emissions
of $\beta ^+$, such that the ACR was efficient as vetos
in catching the two 511~keV photons.
The 80.9~keV $^{133}$Xe line appeared only in Period-III
OFF period. At this low energy, the photo-electric 
cross-section is large compared to the Compton effects.
The residual Compton-scattered photons were of lower energies
and had large absorption by the copper cryostat as well
as by the inactive layer of the HPGe. 
Accordingly,
the effects of this line were confined to the peak region.
The correlated background 
at the low energy signal regions were 
suppressed to 
the level of $\delta _{\pm} <$3e-3.
The short-time-scale fluctuations of 
the unidentified line at 249.8~keV were washed out 
to better than 4\%
when the extended ON/OFF periods were integrated,
giving $\delta _{\pm} <$3e-4.

The stabilities of the ACV and CRV detector systems,
denoted by $\delta$(Detector), 
as well as their effects
on the veto efficiencies [$\delta$(Veto)]
and $\delta _{\pm}$ are summarized in
Table~\ref{bounds}c. 
From Table~\ref{tabcompare},
the raw background rates before the
ACV+CRV cuts were about 
$\phi {\rm (raw)} \sim 30 ~ \cpd$ 
at 30~keV while the
after-cut background levels were
$\phi {\rm (cut)} \sim 1.5 ~ \cpd$,
such that
the combined suppression factors 
were about 5\%,
as shown in Table~\ref{tabselect}.
The fractions of the raw background 
suppressed by each category of selection cuts
are denoted by $f_B$. 
The ON/OFF variations at low-energy are
therefore given by:
\begin{equation}
\delta _{\pm} \sim
[ ~ f_B \cdot \delta {\rm (Veto)} \cdot 
\phi {\rm (raw)} ~ ] ~ / ~ \phi {\rm (cut)} ~~ .
\end{equation}

About 64\% of the raw background were
vetoed by {\it BOTH} ACV and CRV.
These were cosmic-ray events
which penetrated the active ACV and
emitted bremsstrahlung photons which
interacted with the HPGe 
via Compton scatterings.
Besides having the redundancy of signals in
two detector systems, the energy
depositions of these
events were of the order of tens of MeV at
the ACV detectors, where the threshold was
about 5~keV. 
Consequently, the rejection efficiencies
were 100\% and small variations in the
ACV gains or thresholds had negligible
effects on the stabilities.
About 30\% of the raw background were
tagged {\it ONLY} by the ACV detectors. These events
were due to ambient radioactivity and typically
deposited 100~keV to 1~MeV energy at the ACV.
Studies were performed 
with both simulated and real data
on the effects of
the changes in software threshold. 
They indicated that  
an instability in the ACV detector gain of 
$\delta$(Detector)$\sim$5\% 
would lead only to 
a variation in the rejection efficiencies of
$\delta$(Veto)$<$7e-4,
which translated to 
$\delta _{\pm} < $4e-3.
Similarly, only about
1\% of the raw background produced
CRV tag alone. 
These were high-energy bremsstrahlung 
events induced by cosmic-ray interactions
with the passive shielding materials, where
the photons penetrated the ACV system.
Since the PMTs of the CRV system 
were optimized at
the stable plateau region, a 10\% change in the PMT
gain did not give rise to observable change 
in tagging efficiencies
(that is, $\delta$(Veto)$<$1\%), so that
$\delta _{\pm} < $2e-3.
Finally,
about 5\% of the raw background survived 
both ACV+CRV cuts. 
The time variations 
of their survival probabilities 
were accurately measured
to $\delta _{\pm} <$5e-4
by studying the response to
the same software procedures with the
RT events.

\begin{table}[hbt]
\caption{\label{syserror}
The various sources of the systematic
uncertainties, their magnitudes $\delta$(Source)
and their contributions to
$\delta ( \ke10 ^ 2 )$.
In particular, the first entry followed 
from Tables~\ref{stabmonitor}\&\ref{bounds}
and is the dominant contribution.
}
\begin{ruledtabular}
\begin{tabular}{lcr}
Sources & $\delta$(Source) & $\delta ( \ke10 ^ 2 )$ \\ \hline
ON/OFF instabilities [$\delta _{\pm}$(combined)] &  $<$1\% &  $<$0.18 \\
Efficiencies for neutrino events & $<$5e-4 & $<$5e-4 \\
Rates for $\Phi _e ( \munu )$ & $<$3\%  & $<$0.03 \\
$\Phi _e (SM)$ background subtraction & 23\% & 0.03 \\  \hline
Combined Systematic Error & $-$  & $<$0.18 \\
\end{tabular}
\end{ruledtabular}
\end{table}

Taking into account 
the  measured stability levels 
in Table~\ref{stabmonitor}
at the 30~keV signal region
($\delta _{tot} \sim$9e-3),
as well as the combined
upper bounds on the
possible Reactor ON/OFF variations 
($\delta _{\pm} <$7e-3)
in Table~\ref{bounds}(d),
an  uncertainty of 
$\delta _{\pm} {\rm (combined)} <  1\%$ was adopted
to quantify the possible Reactor ON/OFF 
systematic instabilities at the low
energy signal region.
The influence of this 
and other sources of systematic effects 
are summarized in Table~\ref{syserror},
where
the stabilities of the sources are
given by $\delta$(Source) while
their effects on the magnetic moment
searches are represented by
$\delta ( \ke10 ^2 )$.
The stabilities of the
efficiencies of possible 
neutrino-induced signals
in Table~\ref{tabselect}
were accurately measured 
to $\delta$(Source)$<$5e-4 through
the survival probabilities of
the RT events over the various
selection criteria. 
The $\Phi _e ( \munu )$ rates at 10-100~keV depend mostly
on the {\it total} $\nuebar$ flux which is well-modeled to 
$<$2\% at steady-state operation~\cite{bintflux},
as discussed in details in Section~IV.
Residual uncertainties arise from the 
finite rise and fall time of the neutrino flux~\cite{lenu} 
relative to the sharp Reactor ON/OFF instants,
introduced through several long-lived isotopes 
(like $^{93}$Y, $^{97}$Zr, $^{132}$Te) where the
total fission yield is about 15\%.
About 10\% of the data taking time may 
be subjected to this effect, such that
a combined uncertainty of 
$\delta$(Source)$<$3\% was derived
for the evaluation  $\Phi _e ( \munu )$.
A conservative estimate of 30\%
systematic uncertainties for the
$\rnusp$ spectrum below 2~MeV translates
to 23\% uncertainties in the
$\Phi _e (SM)$ background rates below 100~keV. 
However, this only have little impact to the 
accuracies in $\ke10 ^2$ since the effect being
studied in $\Phi _e ( \munu )$ is
an order of magnitude larger than $\Phi _e (SM)$ 
at 10~keV.

\begin{figure}[hbt]
\includegraphics[width=8cm]{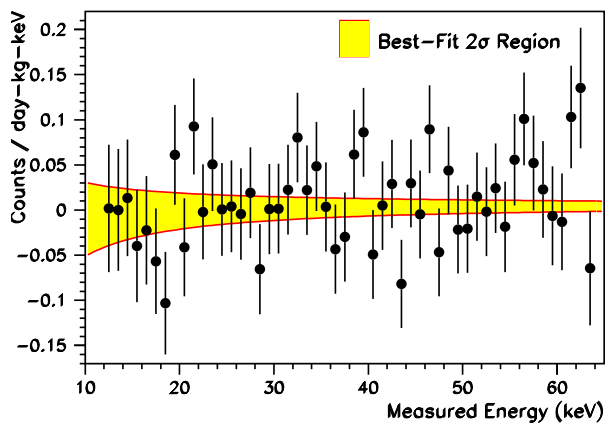}
\caption{
The residual plot 
on the Reactor ON data 
of all periods combined
over the background spectra.
}
\label{residual}
\end{figure}

Combining statistically
the best-fit values on $\ke10 ^2$ for 
the three periods 
and adopting the systematic uncertainties
derived  in Table~\ref{syserror},
the results
\begin{equation}
\rm{
\ke10 ^2 ~ = ~  -0.53 ~ \pm ~ 0.59 (stat.) ~ \pm ~ 0.18 (sys.)
}
\end{equation}
were obtained.
Adopting the unified approach~\cite{pdg04},
a direct limit  on the $\nuebar$ magnetic moment
\begin{equation}
\label{eq::limit}
\rm{
\munuebar  ~ <  ~ 7.4  \times 10^{-11} ~ \mub
}
\end{equation}
at 90\% CL was derived.
The combined residual spectrum for the
three periods of the Reactor 
ON data over the background profiles
is depicted in Figure~\ref{residual}.
The best-fit 2$\sigma$ region for $\rm{ \ke10 ^2}$
is superimposed.
It has been verified that
the correct positive signals 
could be reconstructed by the same
analysis procedures
operating on simulated spectra with 
$\munu$-induced events convoluted with
similar background profiles.
The results are also not sensitive
to alternative choices of $\delta _{\pm}$(combined) 
in Table~\ref{syserror} given the constraints
of the values in 
Tables~\ref{stabmonitor}\&\ref{bounds}.

\begin{figure}[hbt]
\includegraphics[width=8cm]{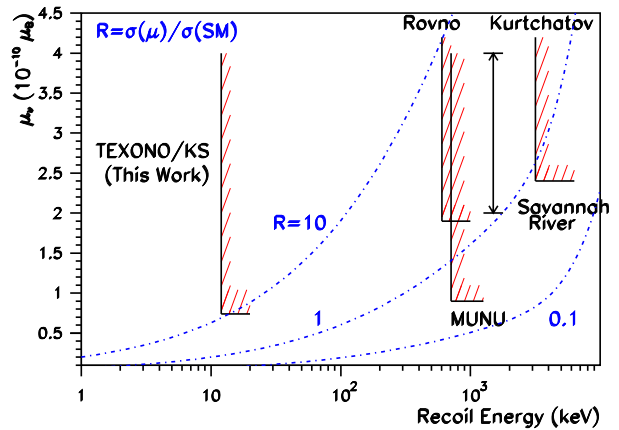}
\caption{
Summary of the results in
the searches of neutrino
magnetic moments with reactor neutrinos.
Both the limits and
the detection thresholds of 
the various experiments are shown. 
}
\label{munusummary}
\end{figure}

Depicted in Figure~\ref{munusummary} is the
summary of the results in $\rm{\munuebar}$ searches
versus the achieved threshold
in reactor experiments.
The dotted lines denote the
ratios between $\munu$-induced and
SM cross-sections
[$\rm{R = \sigma ({\mu}) / \sigma (SM) }$] as
functions of $\rm{ ( T , \munuebar  )}$.
The KS(Ge) experiment has
a much lower threshold of 12~keV compared
to the other measurements.
The large R-values
imply that the KS results
are robust against the uncertainties in the
SM cross-sections. In particular, in
the case where the excess events reported
in Ref.~\cite{vogelengel} 
are due to unaccounted sources of neutrinos, the
limits remain valid.

\begin{figure}[hbt]
\includegraphics[width=8cm]{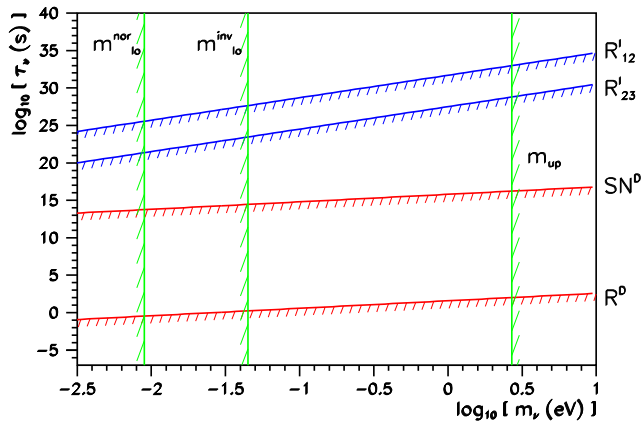}
\caption{
Summary of the excluded parameter
space on
neutrino radiative decay
lifetimes from $\nuebar$ measurements.
Bounds on
$\nu_1$ and $\nu_2$
from reactor
and supernova SN1987a data are
denoted by
R and SN, respectively.
The superscripts  D and I  correspond  to
direct limits and 
indirect bounds inferred from
magnetic moment limits, respectively. 
The subscript ``12'' is attributed
to decays driven by $\rm{\Delta m_{12}^2}$,
and so on.
The upper bound ($\rm{m_{up}}$)
on $\rm{m_{\nu}}$ is due to limits
from direct mass measurements, while
the lower bounds
$\rm{m^{nor}_{lo}}$ and $\rm{m^{inv}_{lo}}$
are valid for
the normal and inverted hierarchies,
respectively.
Bounds for $\rm{R^I_{13}}$ and $\rm{R^I_{23}}$
are represented
by the same bands in this scale.
}
\label{rdksummary}
\end{figure}

Results from oscillation experiments~\cite{pdg04,nu04}
indicate that $\nue$ is predominantly a linear
combination of mass eigenstates $\nu_1$ and $\nu_2$
with mixing angle $\theta_{12}$ given by
$\rm{sin ^2 \theta_{12} \sim 0.27}$.
The mass differences between the mass eigenstates
are
$\rm{\Delta m_{12} ^2 \sim 8 \times 10^{-5} ~ eV^2}$
and
$\rm{\Delta m_{23} ^2 \sim 2 \times 10^{-3} ~ eV^2}$.
Both ``normal'' ($nor$: $\rm{m_3 \gg  m_2 > m_1}$) and
``inverted'' ($inv$:  $\rm{m_2 > m_1 \gg  m_3}$)
mass hierarchies are allowed.
The $\nu_1 \rightarrow \nu_3$ and
$\nu_2 \rightarrow \nu_3$ 
radiative decays
are allowed only in the inverted
mass hierarchy, while
$\nu_2 \rightarrow \nu_1$
is possible in both hierarchies.
Adopting these as input,
the $\munuebar$ limit of
Eq.~\ref{eq::limit} can be translated via
Eq.~\ref{eq::rdk} to  indirect bounds of 
\begin{eqnarray}
\frac{\tau_{13}}{m_{1}^3} ( inv: \nu_1 \rightarrow \nu_3 )
& > & 3.2 \times 10 ^{27}  ~  s / eV^3
\\
\frac{\tau_{23}}{m_{2}^3} ( inv: \nu_2 \rightarrow \nu_3 )
& > & 1.2 \times 10 ^{27}  ~  s / eV^3
\\
\frac{\tau_{21}}{m_{2}^3} ( nor+inv: \nu_2 \rightarrow \nu_1 )
& > & 5.0 \times 10 ^{31}  ~  s / eV^3
\end{eqnarray}
at 90\% CL.
These limits are sensitive to the
bare neutrino-photon
couplings and are therefore
valid for neutrino radiative decays 
in vacuum.
They are summarized 
in Figure~\ref{rdksummary}.
Superimposed are
the limits from the previous direct searches of excess
$\gamma$'s from reactor~\cite{rdkreactor}
and supernova SN1987a~\cite{rdksupernova} neutrinos.
It can be seen that bounds inferred
from $\nu$-e scatterings are much more
stringent than those of the direct approaches.

\section{VIII. Conclusion and Prospects}

This article describes a study of
possible neutrino-photon couplings
using neutrinos from nuclear power reactor
as source.
A germanium detector of target mass
1.06~kg was adopted 
as target where 
data taking and analysis thresholds
of 5~keV and 12~keV, respectively, were achieved.
Sensitive limits were derived 
on neutrino magnetic moments and
neutrino radiative decay lifetimes.
Good background level comparable to those of
the underground dark matter experiments was
achieved at the 10~keV range.
This is the first experiment 
operating with such low threshold
in a high neutrino flux.
The unique data collected with
the HPGe detector at KS open 
the possibilities of
studying other phenomena like
the intrinsic properties
of reactor electron neutrinos~\cite{rnue},
possible neutrino-induced nuclear transitions, and
searches for reactor axions, all of
which are being pursued.

The sensitivities for direct searches of
neutrino magnetic moments~\cite{munureview} 
with neutrino-electron scatterings experiments
scale as
\begin{equation}
\munu \propto \frac{1}{\sqrt{N_{\nu}}} ~ 
[ \frac{B}{M ~  t} ]^{\frac{1}{4}} ~~ ,
\end{equation}
where $N_{\nu}$ is the signal events at some reference magnetic moments,
$B,M,t$ are the background level, detector mass and measurement time,
respectively. The best strategy to
enhance the sensitivities is to 
have an increase on $N_{\nu}$,
which is proportional to the neutrino flux $\phi_{\nu}$ and
is related to the detection threshold.
The atomic energy level effects~\cite{atomic}, however,
limit the potential sensitivity improvement
as the threshold is reduced
below the typical atomic scale of $\sim$1~keV.
There is an on-going experiment GEMMA~\cite{gemma} 
and a proposal MAMONT~\cite{mamont} pursuing along these
directions.

However, this approach cannot improve on the $\munu$ 
sensitivities indefinitely.
Inferring from Figure~\ref{recoil} for reactor
neutrinos, $\Phi _e ( \munu )$
is only $2\%$ of  $\Phi _e ( SM )$ at 1~keV
as $\munu \rightarrow 10 ^{-12} ~ \mub$.
Accordingly, it would be experimentally
difficult to study $\munu$ by looking for anomalous
effects in neutrino-electron scatterings over
SM behaviour for $\munu < 10^{-12} ~ \mub$.
This limitation can be evaded, at least
conceptually, by doing the analog of
``appearance'' experiments in the case
of Majorana neutrinos.
One can look for signatures of anti-neutrinos
of a different flavor 
in a pure and intense neutrino beam which
passes through a dense medium or an strong
magnetic field. Though there is no fundamental
constraint on the lowest reach of the 
detectable $\bar{\nu}$/$\nu$ ratio, 
realistic accelerator-based
experiments are still many orders
of magnitude less sensitive 
than the neutrino-electron scattering
experiments with reactor neutrinos~\cite{simcon}.

The TEXONO Collaboration
meanwhile is pursuing an 
R\&D program on the
``ultra-low-energy'' germanium  detector.
A threshold of 100~eV has been achieved with
a 10~g prototype~\cite{cohsca}.
The goals are to develop a $\sim$1~kg detector
to perform the first
experimental observation of neutrino-nucleus
coherent scattering using reactor neutrinos.
The by-products of such an experiment will
be dark matter searches at the low WIMP mass
regions, as well as the probing of
$\munuebar$ down to
the $\sim 10^{-11} ~ \mub$ range.

\section{IX. Acknowledgments}
 
The experiment reported in this article is the
first particle physics experiment performed
in Taiwan, as well as the first large-scale
scientific collaboration among 
research scientists from Taiwan 
and China~\cite{science03}.
The authors
are indebted to the many colleagues who have
helped to ``make this happen''.
The invaluable contributions by
the technical staff of
our institutes and of the
Kuo-Sheng Nuclear Power Station
are gratefully acknowledged.
The veto scintillator loan from
the CYGNUS~Collaboration is warmly appreciated.
We are also grateful to the referees for
comments on the treatment of systematic
uncertainties.
This work was supported by fundings provided by
the National Science Council and the Academia
Sinica, Taiwan, as well as by the National 
Science Foundation, China.


\begin{thebibliography}{99}

\bibitem{pdg04}
See the respective sections in 
{\it Review of Particle Physics},
Particle Data Group,  
J. Phys. {\bf G 33} (2006),
for details and references.

\bibitem{nu04}
See the respective articles in
{\it Proc. of the XXIst Int. Conf.
on Neutrino Phys. \& Astrophys., 
Paris, France},  
eds. J. Dumarchez, Th. Patzak, and F. Vannucci, 
Nucl. Phys. {\bf B} (Proc. Suppl.) {\bf 143} (2005),
for details and references.

\bibitem{nuprop}
A. de Gouvea, 
Nucl. Phys. {\bf B} (Proc. Suppl.) {\bf 143}, 167 (2005).

\bibitem{prl03}
H.B.~Li et al., 
Phys. Rev. Lett. {\bf 90}, 131802 (2003).

\bibitem{munureview}
H.T.~Wong,
Nucl. Phys. {\bf B} (Procs. Suppl.) {\bf 143}, 205 (2005);
H.T.~Wong and H.B.~Li,
Mod. Phys. Lett. {\bf A 20}, 1103 (2005).

\bibitem{kaysernieves} 
B. Kayser,  Phys. Rev. {\bf D 26}, 1662 (1982); 
J.F. Nieves, Phys. Rev. {\bf D 26}, 3152 (1982).  

\bibitem{munubounds}
N.F.~Bell et al., Phys. Rev. Lett. {\bf 95}, 151802 (2005);
N.F.~Bell et al., Phys. Lett. {\bf B 642}, 377 (2006).

\bibitem{mesm}
B.W. Lee and R.E. Shrock, Phys. Rev. {\bf D 16},
1444 (1977); 
W. Marciano and A.I. Sanda, Phys. Lett. {\bf B 67}, 303 (1977);
K. Fujikawa and R. Shrock, Phys. Rev. Lett. 
{\bf 45}, 963 (1980).

\bibitem{bsm}
R. Shrock, Phys. Rev. {\bf D 9}, 743 (1974);
J. Kim, Phys. Rev. {\bf D 14}, 3000 (1976);
M.A.B. Beg, W.J. Marciano, and M. Ruderman, 
Phys. Rev. {\bf D 17}, 1395 (1977);
M. Fukugita and T. Yanagida, Phys. Rev. Lett. {\bf 58}, 1807 (1987);
S.M. Barr, E.M. Freire, and A. Zee,
Phys. Rev. Lett. {\bf 65}, 2626 (1990).


\bibitem{extradim}
R.N.~Mohapatra, S.P.~Ng, and H.B.~Yu,
Phys. Rev. {\bf D 70}, 057301 (2004).

\bibitem{vogelbeacom}
J.F. Beacom and P. Vogel,
Phys. Rev. Lett. {\bf 83}, 5222 (1999).

\bibitem{sfp}
J. Schechter and J.W.F. Valle, 
Phys. Rev. {\bf D 24}, 1883 (1981).

\bibitem{sfpsolar} 
M.B.~Voloshin, M.I. Vysotskii, and L.B. Okun,
Sov. Phys. JETP {\bf 64}, 446 (1986);
J. Barranco et al., Phys. Rev. {\bf D 66}, 093009 (2002).

\bibitem{kamland}
K. Eguchi et al., Phys. Rev. Lett. {\bf 90}, 021802 (2003);
T. Araki et al.,
Phys. Rev. Lett. {\bf 94}, 081801 (2005).

\bibitem{munusolar}
A.B. Balantekin and C. Volpe,
Phys. Rev. {\bf D 72}, 033008 (2005).

\bibitem{skmunu}
D.W. Liu et al.,
Phys. Rev. Lett. {\bf 93}, 021802 (2004).

\bibitem{raffeltbook}
G.G.~Raffelt, ``Stars as Laboratories for Fundamental
Physics'', Sect. 7.5, U. Chicago Press (1996).

\bibitem{vogelengel}
P.Vogel and J.Engel, Phys. Rev. {\bf D 39}, 3378 (1989).

\bibitem{rdk}
P.P.~Pal and L.~Wolfenstein, 
Phys. Rev. {\bf D 25}, 766 (1982).

\bibitem{rdkmunu}
G.G. Raffelt, Phys. Rev. {\bf D 39}, 2066 (1989).

\bibitem{reines}
F. Reines, H.S. Gurr, and H.W. Sobel,
Phys. Rev. Lett. {\bf 37}, 315 (1976).

\bibitem{kurt}
G.S. Vidyakin et al., JETP Lett. {\bf 55}, 206 (1992).

\bibitem{rovno}
A.I. Derbin et al., JETP Lett. {\bf 57}, 769 (1993).

\bibitem{munu}
C. Amsler et al., Nucl. Instrum. Methods {\bf A 396}, 115 (1997);
Z. Daraktchieva et al., 
Phys. Lett. {\bf B 615}, 153 (2005).

\bibitem{global}
W. Grimus et al., Nucl. Phys. {\bf B 648}, 376 (2003);
M.A.~Tortola, hep-ph/0401135 (2004).

\bibitem{lenu}
H.B.~Li and H.T.~Wong, J. Phys. {\bf G 28}, 1453 (2002).

\bibitem{inersoftware}
FISCOF~Ver 1.0, W.S.~Tong, 08-4-MAN-036-001-1.1, Institute of Nuclear
Energy Research (2001);
FISSRATE~Ver 2.0, W.S.~Kuo,  08-4-MAN-036-002-1.0, Institute of
Nuclear Energy Research (2001).

\bibitem{comsoftware}
CASMO-3~Ver~4.84, Malte~Edenius~et~al., STUDSVIK/SOA-94/9, Studsvik
Scandpower (1994);
SIMULATE-3~Ver~6.07.08, Lorne~Covington~et~al., STUDSVIK/SOA-95/15
Rev.~2, Studsvik Scandpower (2001).

\bibitem{vogel81}
P. Vogel et al., Phys. Rev. {\bf C 24}, 1543 (1981).

\bibitem{ncaprussian}
V.I. Kopeikin, L.A. Mikaelyan, and V.V. Sinev,
Phys. Atom. Nucl. {\bf 60}, 172 (1997).

\bibitem{rnue}
B. Xin et al., Phys. Rev. {\bf D 72}, 012006 (2005).

\bibitem{bintflux}
Y. Declais et al., Phys. Lett. {\bf B 338}, 383 (1994).

\bibitem{bugey3}
B. Achkar et al.,  Phys. Lett. {\bf B 374},
243 (1996).

\bibitem{ksprogram}
H.T.~Wong, Mod. Phys. Lett. {\bf A 19}, 1207 (2004).

\bibitem{kscsi}
H.B.~Li et al.,
Nucl. Instrum. Methods {\bf A 459}, 93 (2001).

\bibitem{eledaq}
W.P.~Lai et al.,
Nucl. Instrum. Methods {\bf A 465}, 550 (2001).

\bibitem{canberra}
ULB-HPGe Model GC5019, Canberra Industries, USA.

\bibitem{vme}
VME-PCI Adaptor, Model 618, SBS Technologies, USA.

\bibitem{raid}
Redundant Array of Independent Disks (RAID), Ultra Trak SX8000,
Promise Technology, USA.

\bibitem{atomic}
V.I. Kopeikin et al., Phys. Atom. Nucl. {\bf 60}, 2032 (1997).

\bibitem{rdkreactor}
L. Oberauer, F. von Feilitzsch, and R.L. M\"{o}ssbauer,
Phys. Lett. {\bf B 198}, 113 (1987); 
J. Bouchez et al., Phys. Lett. {\bf B 207}, 217 (1988). 

\bibitem{rdksupernova}
E.L. Chupp, W.T. Vestrand, and C. Reppin,
Phys. Rev. Lett. {\bf 62}, 505 (1989).

\bibitem{gemma}
A.G.~Beda et al., Phys. Atom. Nucl. {\bf 67}, 
1948 (2004).

\bibitem{mamont}
L.N. Bogdanova, Nucl. Phys. {\bf A 721}, 499 (2003)

\bibitem{simcon}
M.C. Gonzalez-Garcia, F. Vannucci, and J. Castromonte,
Phys. Lett. {\bf B 373}, 153 (1996); 
J.-M.~Fr\`{e}re, R.B.~Nevzorov, and M.I.~Vysotsky,
Phys. Lett. {\bf B 394}, 127 (1977).

\bibitem{cohsca}
H.T.~Wong, J. Phys. Conf. Ser. {\bf 39}, 266 (2006).

\bibitem{science03}
D. Normile, Science {\bf 300}, 1074 (2003).

\end{thebibliography}
\end{document}